\documentclass{cas-sc}

\usepackage[authoryear]{natbib}
\usepackage{amssymb}
\usepackage{amsmath}
\usepackage{upgreek}
\usepackage{caption}
\usepackage{float}
\usepackage{booktabs}
\usepackage{subfig}
\usepackage{setspace}
\usepackage{graphicx} 
\usepackage{float}
\usepackage{algorithm}  
\usepackage{algpseudocode}
\usepackage{color}
\usepackage{setspace}
\def\tsc#1{\csdef{#1}{\textsc{\lowercase{#1}}\xspace}}
\tsc{WGM}
\tsc{QE}
\tsc{EP}
\tsc{PMS}
\tsc{BEC}
\tsc{DE}
\usepackage{lineno}
\begin{document}

\let\WriteBookmarks\relax
\def\floatpagepagefraction{1}
\def\textpagefraction{.001}
\shorttitle{RTD learning from SIP data}
\shortauthors{Bérubé et al.}

\title [mode = title]{Learning relaxation time distributions from spectral induced polarization data with a complex-valued variational autoencoder}

\author[1]{Charles L. Bérubé}[orcid=0000-0002-5438-0382]
\ead{charles.berube@polymtl.ca}

\author[1]{Sébastien Gagnon} 

\author[1]{Lahiru M.A. Nagasingha}

\author[2]{Jean-Luc Gagnon}
\author[1]{E. Rachel Kenko}

\author[3]{Reza Ghanati}

\author[4]{Frédérique Baron}

\address[1]{Civil, geological and mining engineering department, Polytechnique Montréal, C.P. 6079, succ. Centre-ville, Montréal, H3C 3A7, QC, Canada}
\address[2]{Engineering physics department, Polytechnique Montréal, C.P. 6079, succ. Centre-ville, Montréal, H3C 3A7, QC, Canada}
\address[3]{Institute of geophysics, University of Tehran, Tehran, 4359-44411, Iran} 
\address[4]{Physics department, Université de Montréal, C.P. 6128, succ. Centre-ville, Montréal, H3C 3J7, QC, Canada}

\begin{abstract}
Spectral induced polarization (SIP) is a geophysical method used to characterize subsurface materials. It measures the frequency-dependent complex electrical resistivity of rocks and soils through the application of an alternating current. Debye and Warburg decompositions are standard methods for analyzing and interpreting SIP data, as they allow estimation of the relaxation time distribution (RTD) of geomaterials. However, conventional decomposition approaches treat measurements independently, work in real-valued spaces despite the complex-valued nature of SIP data, and provide limited uncertainty quantification. These limitations reduce the effectiveness of conventional decomposition on heterogeneous datasets. We reformulate the decomposition approach as an unsupervised machine learning problem and introduce a conditional variational autoencoder (CVAE) that learns a shared mapping from SIP datasets to continuous RTDs. The model is validated on a dataset comprising 140 laboratory and field SIP measurements of granular mixtures, sedimentary and igneous rock, and cementitious materials. The CVAE operates in complex-valued data space and achieves reconstruction errors of 0.53\,\% and 0.27\,\% for the imaginary and phase components of resistivity, respectively, with evidence of improvement over conventional methods at corresponding $p$-values of $0.007$ and $0.06$. The inferred RTDs are stable and physically consistent, and their total chargeability and mean relaxation time correlate with polarizable grain content and grain size, respectively, with coefficients of determination up to 0.98. Sensitivity analysis shows that the decomposition depends primarily on the placement and weighting of relaxation modes, while global scaling parameters play a minor role. An additional contribution of the proposed method is the learned latent representation, which organizes SIP spectra into a structured space. Unsupervised clustering in a two-dimensional projection of this space improves the silhouette score by nearly a factor of two relative to conventional RTD parameters. These results demonstrate that the CVAE enables accurate and interpretable analysis of SIP data across diverse datasets.
\end{abstract}
 
% \begin{coverletter}

% Dear Editors-in-Chief,
% \newline
 
% please find the enclosed manuscript "..." which we are submitting for exclusive consideration for publication in Computers \& Geosciences. We confirm that the submission follows all the requirements and includes all the items of the submission checklist.  
% \newline
 
% The manuscript presents ... 
% \newline

% We provide the source codes in a public repository with details listed in the section "Code availability".
% \newline

% Thanks for your consideration. 
% \newline

% Sincerely,
% \newline

% Authors names

% Corresponding author affiliation and e-mail
% \newline

% \textbf{Delete before submission:}

% Please confirm that your submission follows all the requirements of the guidelines, including the submission checklist:

% - Cover letter

% - Highlights

% - Authorship statement

% - The manuscript must be single column and double spaced

% - Reference must be in the author-date format

% - Code availability section 

% *The manuscripts that do meet the requirement guidelines will be desk-rejected.

% \end{coverletter}

% \begin{highlights}
% \item Highlight 1
% \item Highlight 2
% \item Highlight 3
% \item Highlight 4
% \item Highlight 5
% \end{highlights}


\maketitle 

\printcredits

\onehalfspacing

\section{Introduction}
\label{sec:intro}

Spectral induced polarization (SIP) is a geophysical method that measures the frequency-dependent, complex-valued electrical resistivity of the subsurface. It aims to characterize the ability of geomaterials to temporarily and reversibly store electrical charges through interfacial polarization processes. These processes are commonly represented by a relaxation time distribution (RTD) estimated with the Debye \citep{morgan_inversion_1994, nordsiek_new_2008, zorin_spectral_2015} or Warburg \citep{revil_spectral_2014} decomposition methods. However, decomposition is an ill-posed inverse problem that requires regularization and parameter choices. Deterministic approaches stabilize RTD estimation using techniques such as L-curve analysis \citep{florsch_inversion_2014}, whereas stochastic formulations use a polynomial approximation of the RTD to enable uncertainty quantification \citep{keery_markov-chain_2012,berube_bayesian_2017}. Newer developments use complex conductivity formulations and extensions to time-lapse and time-domain measurements \citep{ustra_relaxation_2016,weigand_debye_2016,hase_conversion_2023}. 

Chargeability, indicative of polarization intensity, and relaxation time, which depends on the characteristic polarization frequency, are key RTD parameters that relate to rock and soil properties. For example, \citet{weller_estimating_2010} and \citet{weller_estimation_2010} relate chargeability to the specific surface area normalized by pore volume in sands and sandstones. Additionally, \citet{zisser_dependence_2010} and \citet{bairlein_influence_2014} document the influence of temperature and water saturation on RTD parameters. Variations of the RTD have also been observed during desaturation processes \citep{martin_desaturation_2021} and as a function of grain surface characteristics \citep{zibulski_influence_2023}. Numerous studies thus leverage SIP decomposition to investigate hydraulic conductivity and pore characteristics in subsurface materials \citep{attwa_spectral_2013, weller_relationship_2013, revil_spectral_2014, ustra_spectral_2012}, contamination and geomaterial degradation processes \citep{flores_orozco_delineation_2012, khajehnouri_non-destructive_2020}, and mineral composition, grain size, and alteration intensity \citep{gurin_time_2013, gurin_application_2015, berube_mineralogical_2019}.

Despite their widespread use, SIP decomposition methods have computational limitations. First, most approaches require relaxation time discretization \citep[e.g.,][]{nordsiek_new_2008}, which imposes a prior structure on the RTD and makes the results sensitive to smoothing weights, grid spacing, and initialization \citep{weigand_debye_2016}. While polynomial RTD parameterizations may improve numerical stability, they also introduce additional hyperparameters that influence the solution \citep[e.g.,][]{keery_markov-chain_2012}. Second, conventional decomposition treats each spectrum independently and does not exploit shared structure across datasets, which limits scalability and prevents learning of population-level representations. Third, existing decomposition methods operate in real-valued spaces, despite SIP data being complex-valued, which may hinder SIP parameter estimation \citep{berube_complex-valued_2025}.

We propose a complex-valued conditional variational autoencoder (CVAE) that learns continuous RTDs and their uncertainties without predefined discretization or polynomial parameterization. The approach replaces individual spectrum inversion with a shared mapping learned from multiple spectra. Moreover, the proposed formulation differs from standard CVAE implementations because it embeds a relaxation model in the decoder and operates in complex-valued data spaces. The objectives are to (1) learn continuous RTD representations from SIP data, (2) characterize the latent space for dataset-level representation and clustering, and (3) evaluate reconstruction accuracy and parameter interpretability relative to conventional decomposition. We first introduce the RTD formulation and describe the CVAE model selection process. We then quantify reconstruction accuracy on laboratory and field data, evaluate the consistency of the inferred RTDs, and analyze the model's sensitivity, latent space structure, and generative capabilities before comparing the proposed approach with conventional decomposition.

\section{Theory}
\label{sec:theory}

\subsection{Forward model}
We formulate the forward model in terms of conductivity throughout this work, although SIP measurements, including those presented here, are commonly reported as resistivity \citep{macnae_physical_2025, revil_comment_2026}. Following \citet{florsch_inversion_2014}, the modelled complex conductivity is
\begin{equation}
\hat{\sigma}(\omega) = \sigma_\infty - M \int_{0}^{\infty} \Gamma(\tau)\, \lambda(\omega,\tau)\, \mathrm{d}\tau + i\omega\varepsilon_\infty, 
\end{equation}
where $i^2=-1$, $\omega$ is the angular frequency in rad/s, $\tau$ is the relaxation time in s, $\sigma_\infty$ is the instantaneous conductivity in S/m, and $M=\sigma_\infty-\sigma_0\geq0$ is polarization amplitude, with $\sigma_0$ denoting the direct current conductivity. The parameter $\varepsilon_\infty$ is the high-frequency permittivity in F/m. The RTD, $\Gamma(\tau)$, is normalized such that 
\begin{equation}
    \int_{0}^{\infty} \Gamma(\tau)\,\mathrm{d}\tau=1,
\end{equation} 
and the relaxation kernel is
\begin{equation}
   \lambda(\omega, \tau)=\frac{1}{1+(i\omega\tau)^c},
\end{equation}
where $c=1$ in the case of Debye decomposition. Using $c=0.5$ leads to Warburg decomposition \citep{revil_spectral_2014} and better represents the polarization of the Stern layer and that of a leaking capacitance. 

Using the dimensionless chargeability \citep{seigel_mathematical_1959}
\begin{equation}
    m=\frac{\sigma_\infty-\sigma_0}{\sigma_\infty},
\end{equation}
we rewrite the modelled conductivity as
\begin{equation}
\hat{\sigma}(\omega) = \sigma_\infty \left[ 1 - m \int_{0}^{\infty} \Gamma(\tau)\, \lambda(\omega,\tau)\, \mathrm{d}\tau \right] + i\omega\varepsilon_\infty. \label{eq:cond_continuous}
\end{equation}
We also express the RTD in logarithmic time by using $u=\ln\tau$ and defining $\gamma(u) = \tau\,\Gamma(\tau)$. This quantity satisfies
\begin{equation}
\int_{-\infty}^{\infty} \gamma(u)\,\mathrm{d}u = 1,
\label{eq:rtd_norm}
\end{equation}
and Eq.~\eqref{eq:cond_continuous} becomes
\begin{equation}
\hat{\sigma}(\omega) = \sigma_\infty \left[ 1 - m \int_{-\infty}^{\infty} \gamma(u)\, \lambda(\omega,e^u)\, \mathrm{d}u \right] + i\omega\varepsilon_\infty. 
\label{eq:cond_log_continuous} \end{equation}
For comparison with complex resistivity data, we calculate the modelled resistivity from the conductivity response as
\begin{equation}
\hat{\rho}(\omega)
=
\frac{1}{\hat{\sigma}(\omega)}.
\label{eq:res_log_continuous}
\end{equation}

\subsection{Integrating parameters}
Peak relaxation times, total chargeability, and mean relaxation time are used for RTD interpretation. Peak relaxation times $\tau^{\ast}$ are the local maxima of $\gamma(u)$ with $u^{\ast}=\ln\tau^\ast$. Total chargeability truncated over the interval $[\tau_1,\tau_2]$ is
\begin{equation}
m_{[\tau_1,\tau_2]}
=
m\int_{u_1}^{u_2}
\gamma(u)\,\mathrm{d}u.
\end{equation}
The subscript $[\tau_1,\tau_2]$ denotes the relaxation time interval over which chargeability is integrated. Because $\gamma(u)$ is normalized and $0<m<1$, the truncated chargeability satisfies $0\leq m_{[\tau_1,\tau_2]}\leq m<1$. The truncated mean relaxation time, $\bar{\tau}=\exp(\bar{u})$, is obtained with
\begin{equation}
\bar{u}_{[u_1,u_2]}
=
\frac{
\int_{u_1}^{u_2}
u\,\gamma(u)\,\mathrm{d}u
}{
\int_{u_1}^{u_2}
\gamma(u)\,\mathrm{d}u
}.
\end{equation}

%%% METHODS %%%
\section{Methods}

\subsection{Data preprocessing}

Let an SIP dataset with $J$ spectra measured at $K$ frequencies be defined as
\begin{equation}
\mathcal{S}
=
\left\{
\left(f_k,\, \rho_{j,k},\, \Delta\rho_{j,k}\right)
\;\middle|\;
j=1,\dots,J,\; k=1,\dots,K
\right\},
\end{equation}
where $f_k$ is the $k$\textsuperscript{th} measurement frequency in~Hz,
$\rho_{j,k}$ is the measured complex resistivity of sample $j$,
and $\Delta\rho_{j,k}$ is its associated uncertainty. We write the complex resistivity as
\begin{equation}
\rho_{j,k}
=
|\rho_{j,k}|e^{i\varphi_{j,k}}
=
\rho_{j,k}' + i\rho_{j,k}'',
\end{equation}
where $\varphi_{j,k}$ is the phase shift in rad, and where
$\rho_{j,k}'$ and $\rho_{j,k}''$ denote the real and imaginary parts of resistivity
in $\Omega\,\mathrm{m}$, respectively. Each SIP spectrum is represented by
$
{\boldsymbol{\rho}}_j
=
[\rho_{j,1},\dots,\rho_{j,K}]
\in \mathbb{C}^{K}
$,
with uncertainty vector
$
\Delta{\boldsymbol{\rho}}_j
=
[\Delta\rho_{j,1},\dots,\Delta\rho_{j,K}]
\in \mathbb{C}^{K}
$
and common frequency vector
$
{\boldsymbol{f}}
=
[f_1,\dots,f_K]
\in \mathbb{R}^{K}
$.

We normalize each spectrum by the resistivity amplitude at the lowest frequency,
$|\rho(f_{\min})|$, such that
\begin{equation}
\tilde{\rho}_k = \frac{\rho_k}{|\rho(f_{\min})|},
\qquad
\Delta\tilde{\rho}_k = \frac{\Delta\rho_k}{|\rho(f_{\min})|}.
\label{eq:normalization}
\end{equation}
This normalization constrains the data to a compact region of the complex plane. The preprocessed dataset is thus
\begin{equation}
\mathcal{\tilde{S}}
=
\left\{
\left(f_k,\, \tilde{\rho}_{j,k},\, \Delta\tilde{\rho}_{j,k}\right)\right\}.
\end{equation}

\subsection{Conditional variational autoencoder}

\begin{figure}
\centering
\includegraphics[width=1.0\textwidth]{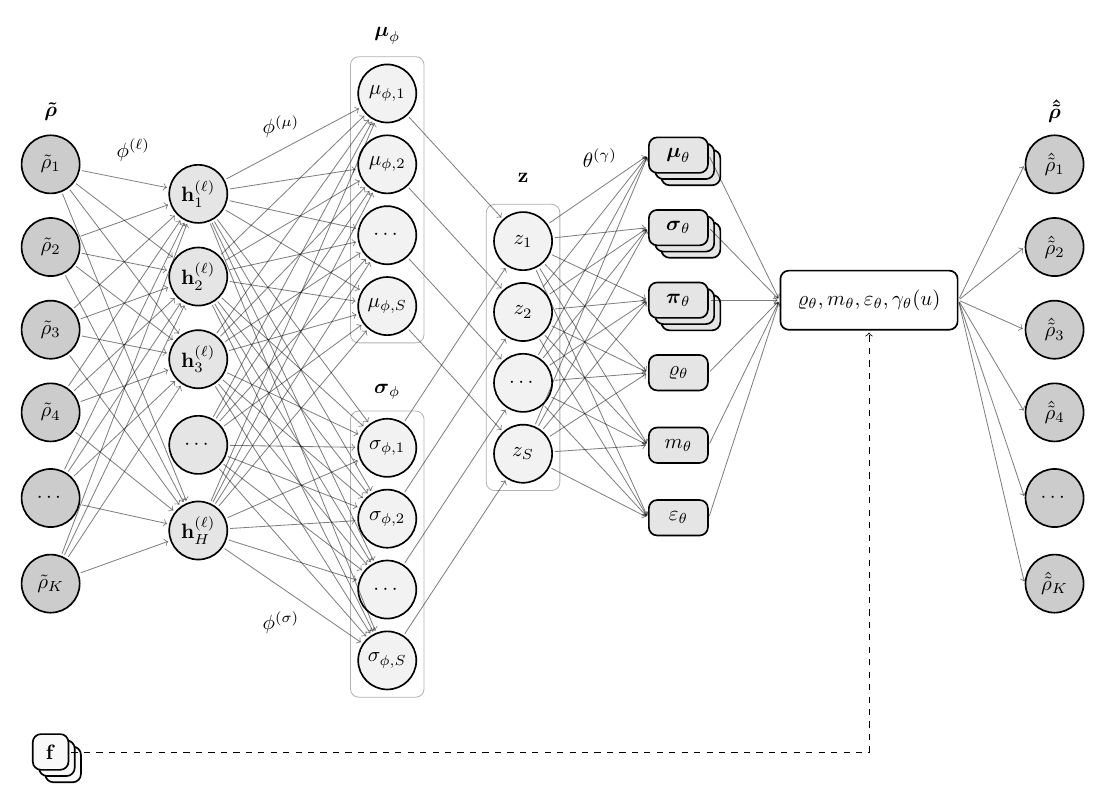}
\caption{
Architecture of the CVAE used for decomposition. The encoder maps the normalized complex resistivity $\boldsymbol{\tilde{\rho}}$ to the parameters $\boldsymbol{\mu}_\phi$ and $\boldsymbol{\sigma}_\phi$ of a
diagonal Gaussian distribution. Latent samples $\mathbf{z}$ are transformed by the decoder, conditioned on the
measurement frequencies $\mathbf{f}$, into the parameters of a Gaussian mixture representation of the logarithmic RTD $\gamma_\theta(u)$. The RTD feeds the forward model in Eq.~\eqref{eq:cvae_output} to reconstruct the normalized complex resistivity $\boldsymbol{\hat{\tilde{\rho}}}$.
}
\label{fig:cvae}
\end{figure}

The nonlinear relationship between SIP data and a low-dimensional latent representation is modelled using a CVAE \citep{sohn_learning_2015}. The CVAE follows the sequence
\begin{equation}
\boldsymbol{\tilde{\rho}}
\;\xrightarrow{\mathrm{encode}}\;
\left(\boldsymbol{\mu}_{\phi},\ln\boldsymbol{\sigma}^{2}_{\phi}\right)
\;\xrightarrow{\mathrm{reparameterize}}\;
\mathbf{z}
\;\xrightarrow{\mathrm{decode}\mid\mathbf{f}}\;
\left(\gamma_\theta,\,\boldsymbol{\hat{\tilde{\rho}}}\right),
\end{equation}
where the encoder, with trainable parameters $\phi$, maps the normalized resistivity $\boldsymbol{\tilde{\rho}}$ to the mean $\boldsymbol{\mu}_{\phi}$ and log-variance $\ln\boldsymbol{\sigma}^2_{\phi}$ of a diagonal Gaussian approximate posterior. The decoder, with trainable parameters $\theta$ and conditioning frequencies $\mathbf{f}$, maps latent distribution samples $\mathbf{z}$ to the RTD $\gamma_\theta$ and its associated resistivity model $\boldsymbol{\hat{\tilde{\rho}}}$. We implement the CVAE, shown in Figure~\ref{fig:cvae}, using PyTorch \citep{paszke_pytorch_2019}.

%TC:ignore
%TC:endignore

\subsubsection{Encoder}

The encoder defines a mapping $\mathcal{E}_\phi$ from $\boldsymbol{\tilde{\rho}}$ to the parameters of a diagonal Gaussian latent distribution, i.e.,
\begin{equation}
\left(\boldsymbol{\mu}_\phi,\;
\ln\boldsymbol{\sigma}_\phi^{2}\right)
=
\mathcal{E}_\phi(\boldsymbol{\tilde{\rho}}).
\end{equation}
The encoder uses a sequence of complex-valued affine transformation layers, each followed by an activation function $\psi$. Starting from the input layer $\mathbf{h}_0=\boldsymbol{\tilde{\rho}}$, the hidden representations evolve as
\begin{equation}
\mathbf{h}_{\ell+1}
=
\psi\!\left(
\mathbf{W}_{\ell}\mathbf{h}_{\ell} + \mathbf{b}_{\ell}
\right),
\qquad \ell=0,\dots,L-1,
\end{equation}
where $\mathbf{W}_{\ell}\in\mathbb{C}$ and $\mathbf{b}_{\ell}\in\mathbb{C}$ denote the weight matrices and bias vectors of the $\ell$\textsuperscript{th} encoder layer. After $L$ layers, the encoder yields a final hidden representation $\mathbf{h}_L\in\mathbb{C}^{H}$, where $H$ is the constant width of all hidden layers. For $\psi$, we use the complex cardioid \citep{virtue_better_2017}, reading
\begin{equation}
\psi(x)
=
\frac{x\left(1+\cos(\arg x)\right)}{2},
\qquad x\in\mathbb{C},
\end{equation}
because it is effective for SIP data \citep{berube_complex-valued_2025}.

Two parallel transformations map the final hidden representation
$\mathbf{h}_L$ to 
\begin{equation}
\boldsymbol{\mu}_{\phi}
=
\Re\!\left(\mathbf{W}_{\mu}\mathbf{h}_{L} + \mathbf{b}_{\mu}\right)
+
\Im\!\left(\mathbf{W}_{\mu}\mathbf{h}_{L} + \mathbf{b}_{\mu}\right),
\end{equation}
and
\begin{equation}
\ln\boldsymbol{\sigma}_{\phi}^{2}
=
\Re\!\left(\mathbf{W}_{\sigma}\mathbf{h}_{L} + \mathbf{b}_{\sigma}\right)
+
\Im\!\left(\mathbf{W}_{\sigma}\mathbf{h}_{L} + \mathbf{b}_{\sigma}\right).
\end{equation}
These quantities define the Gaussian approximate posterior
\begin{equation}
q_{\phi}\left(\mathbf{z}\mid\boldsymbol{\tilde{\rho}}\right)
=
\mathcal{N}\!\left(
\boldsymbol{\mu}_{\phi},
\;\operatorname{diag}\!\left(
\boldsymbol{\sigma}_{\phi}^{2}
\right)
\right), \qquad \mathbf{z}\in\mathbb{R}^S,
\label{eq:posterior}
\end{equation}
where $S$ denotes the latent space dimensionality.

\subsubsection{Latent representation}
The latent variables are sampled using the method of
\cite{kingma_auto-encoding_2014}. Given the encoder outputs
$\boldsymbol{\mu}_{\phi}$ and $\ln\boldsymbol{\sigma}_{\phi}^{2}$, we draw
$\boldsymbol{\epsilon}\sim\mathcal{N}(\mathbf{0},\mathbf{I}_{S})$, where
$\mathbf{I}_{S}$ is the $S\times S$ identity matrix, and compute
\begin{equation}
\mathbf{z}
=
\boldsymbol{\mu}_{\phi}
+
\boldsymbol{\sigma}_{\phi}
\odot
\boldsymbol{\epsilon},
\label{eq:reparametrization}
\end{equation}
where $\odot$ denotes the Hadamard product.

\subsubsection{Decoder}

The decoder maps $\mathbf{z}$ to $\boldsymbol{\hat{\tilde{\rho}}}$, given $\mathbf{f}$, through a logarithmic RTD parameterized as a Gaussian mixture. First, the decoder defines a mapping $\mathcal{D}_\theta$ from $\mathbf{z}$ to six groups of raw parameters, 
\begin{equation} \left(\tilde{\varrho}_\theta,\; \tilde{m}_\theta,\; \tilde{\varepsilon}_\theta,\; \boldsymbol{\tilde{\pi}}_{\theta},\; \boldsymbol{\tilde{\mu}}_{\theta},\; \boldsymbol{\tilde{\sigma}}_{\theta}\right) = \mathcal{D}_\theta(\mathbf{z}\,|\,\mathbf{f}), 
\end{equation} 
where $\boldsymbol{\tilde{\pi}}_{\theta}, \boldsymbol{\tilde{\mu}}_{\theta}, \boldsymbol{\tilde{\sigma}}_{\theta}\in\mathbb{R}^{N}$, with $N$ the number of Gaussian mixture components, and where $\tilde{\varrho}_\theta,\tilde{m}_\theta, \tilde{\varepsilon}_\theta\in\mathbb{R}$.

The raw decoder outputs are transformed into physically admissible RTD parameters. Scale parameters are mapped to strictly positive values, whereas those with physical ranges are mapped to their bounded intervals. Thus, the direct current resistivity $\rho_0$ is estimated through
\begin{equation}
\varrho_\theta
=
\left(1 + 0.1\,\tanh(\tilde{\varrho}_\theta)\right),
\qquad
0.9 \le \varrho_\theta \le 1.1,
\end{equation}
assuming that $\rho_0 = \varrho_\theta\times|\rho(f_{\min})|$. Dimensionless chargeability is predicted by the logistic function,
\begin{equation}
m_\theta = \frac{1}{1+\exp(-\tilde{m}_\theta)}, \qquad 0 < m_\theta < 1,
\end{equation}
and the normalized high-frequency permittivity parameter reads
\begin{equation}
    \varepsilon_{\theta}
    =
    \exp\!\left(\tilde{\varepsilon}_{\theta}\right), \qquad \varepsilon_{\theta} > 0, 
\end{equation}
which is related to the physical high-frequency permittivity through
\begin{equation} \varepsilon_\theta = \varepsilon_{\infty} \left|\rho(f_{\min})\right|. \end{equation} 
Since $\rho_0=\varrho_\theta\times\left|\rho(f_{\min})\right|$, it follows that 
\begin{equation} \varepsilon_\theta\varrho_\theta = \varepsilon_{\infty}\rho_0. 
\end{equation}

The Gaussian mixture weights use the softmax function,
\begin{equation}
\pi_{\theta,n}
=
\frac{\exp(\tilde{\pi}_{\theta,n})}
{\sum_{l}\exp(\tilde{\pi}_{\theta,l})},
\qquad n=1,\dots,N,
\end{equation}
and the decoder places each mixture component at a relaxation time
$\tau = e^{u}$. Using $\omega_k=2\pi f_k$, the bounds for $u$ are derived from $\omega_k^{-1}$ in common logarithmic space, expanded by two decades, and converted to natural logarithmic coordinates as
\begin{equation}
u_{\min}
=
\ln10\times
\left(
\left\lfloor 
\min\nolimits_{k}\;\log_{10}\!\big(\omega_k^{-1}\big)
\right\rfloor
-2
\right),
\label{eq:bounds1}
\end{equation}
and
\begin{equation}
u_{\max}
=
\ln10\times
\left(
\left\lceil 
\max\nolimits_{k}\;\log_{10}\!\big(\omega_k^{-1}\big)
\right\rceil
+2
\right),
\label{eq:bounds2}
\end{equation}
where $\lfloor\cdot\rfloor$ and $\lceil\cdot\rceil$ denote the floor and ceiling operators, respectively. 

For numerical evaluation, the decoder represents the RTD as a finite Gaussian mixture. Let $u_{r}\in[u_{\min},u_{\max}]$ denote a quadrature grid with spacing
$\Delta u$.
The means of the Gaussian mixture components follow
\begin{equation}
    \mu_{\theta,n}
    =
    u_{\min}
    +
    \frac{u_{\max}-u_{\min}}{2}\,
    \big[\tanh(\tilde{\mu}_{\theta,n})+1\big],
    \qquad n=1,\dots,N,
\end{equation}
and each Gaussian width is strictly positive through
\begin{equation}
    \sigma_{\theta,n}
    =
    \exp\!\left(\tfrac12\,\tilde{\sigma}_{\theta,n}\right),
    \qquad n=1,\dots,N.
\end{equation}
The raw logarithmic RTD thus reads
\begin{equation}
\tilde{\gamma}_\theta(u_{r})
=
\sum_{n=1}^{N}
\pi_{\theta,n}\,
\exp\!\left[
    -\frac{(u_{r}-\mu_{\theta,n})^{2}}{2\sigma_{\theta,n}^{2}}
\right],
\end{equation}
where the mixture weights satisfy $\pi_{\theta,n}\ge0$ and $\sum_{n=1}^{N}\pi_{\theta,n}=1$. To satisfy the discrete analogue of Eq.~\eqref{eq:rtd_norm}, we normalize the RTD on the quadrature grid as
\begin{equation}
\gamma_\theta(u_{r})
=
\frac{
    \tilde{\gamma}_\theta(u_{r})
}{
    \sum_{r}
    \tilde{\gamma}_\theta(u_{r})\,\Delta u
}.
\end{equation}

Using $\sigma_0=(1-m_\theta)\sigma_\infty$ and $\varepsilon_\theta\varrho_\theta = \varepsilon_{\infty}\rho_0$, the RTD integral in Eq.~\eqref{eq:cond_log_continuous} is evaluated by numerical quadrature as
\begin{equation} 
\frac{\hat{\sigma}(\omega_k)}{\sigma_0} = \frac{ 1 - m_\theta \sum_{r} \gamma_\theta(u_r)\, \lambda(\omega_k,e^{u_r})\, \Delta u }{ 1-m_\theta } + i\omega_k \varepsilon_\theta \varrho_\theta\label{eq:decoder_cond}
\end{equation}
and the modelled complex resistivity is
\begin{equation} 
\frac{\hat{\rho}(\omega_k)}{\rho_0} = \left[ \frac{\hat{\sigma}(\omega_k)}{\sigma_0}\right]^{-1}. \label{eq:decoder_res}
\end{equation}
Finally, the decoder output is normalized to match the convention used in Eq.~\eqref{eq:normalization}, i.e.,
\begin{equation}
\hat{\tilde{\rho}}(\omega_k)
=
\varrho_\theta
\frac{\hat{\rho}(\omega_k)}{\rho_0}.
\label{eq:cvae_output}
\end{equation}

\subsection{Relaxation kernel choice}
The recovered RTD depends on the selected relaxation kernel, and comparisons among datasets therefore require a consistent kernel and decomposition procedure \citep{weigand_relationship_2016}. We use a Debye kernel with $c=1$. This choice is motivated by four considerations: (1) the proposed method is evaluated against conventional Debye decomposition methods; (2) mechanistic models of metallic particle polarization predict Debye-like relaxation in the absence of active cations \citep{wong_electrochemical_1979, misra_interfacial_2016, bucker_electrochemical_2018}; (3) the investigated materials mostly contain metallic, semimetallic, or semiconducting inclusions; and (4) tests show that the Debye kernel better reconstructs the SIP spectra analyzed in this study. Nevertheless, the proposed method is not restricted to $c=1$. The accompanying code supports Debye, Warburg, and user-defined exponents. An optimal $c$ value can also be inferred by the decoder.

\subsection{Neural network optimization}

We aim to maximize the evidence lower bound (ELBO), defined for each measured spectrum as
\begin{equation}
\operatorname{ELBO}\left(\theta,\phi;\boldsymbol{\tilde{\rho}},\mathbf{f}\right)
=
\mathbb{E}_{q_{\phi}\left(\mathbf{z}\,|\,\boldsymbol{\tilde{\rho}}\right)}
\!\left[
\ln p_{\theta}\left(\boldsymbol{\tilde{\rho}}\,|\,\mathbf{z},\mathbf{f}\right)
\right]
-
D_{\mathrm{KL}}\!\Big(
q_{\phi}(\mathbf{z}\,|\,\boldsymbol{\tilde{\rho}})
\;\big\|\;
p(\mathbf{z})
\Big),
\label{eq:elbo}
\end{equation}
where $D_\mathrm{KL}$ denotes the Kullback--Leibler divergence between the approximate posterior of $\mathbf{z}$ and its prior.

\subsubsection{Negative log-likelihood}

The first term in Eq.~\eqref{eq:elbo} evaluates the expected log-likelihood of $\boldsymbol{\tilde{\rho}}$
under the decoder prediction
$\boldsymbol{\hat{\tilde{\rho}}}$.
The residual vector is
\begin{equation}
\boldsymbol{\eta}
=
\boldsymbol{\tilde{\rho}}
-
\boldsymbol{\hat{\tilde{\rho}}}
=
(\eta_{1},\ldots,\eta_{K}),
\qquad
\eta_{k}=\eta_{k}'+i\,\eta_{k}'',
\end{equation}
and a Gaussian likelihood is adopted for each
frequency. Conditioned on $\mathbf{z}$ and $\mathbf{f}$,
the decoder defines the likelihood
\begin{equation}
p_{\theta}\!\left(\boldsymbol{\tilde{\rho}}\mid\mathbf{z},\mathbf{f}\right)
=
\prod_{k=1}^{K}
p\!\left(
\eta_k;\, v_k,\, \delta_k
\right),
\end{equation}
where the variance $v_k$ and pseudo-variance $\delta_k$
characterize the second-order statistics of the residual. The corresponding density is derived in Appendix~A and reads
\begin{equation}
p(\eta_k; v_k, \delta_k)
=
\frac{1}{\pi\sqrt{\,v_k^{2}-|\delta_k|^{2}\,}}
\exp\!\left(
-\frac{
v_k|\eta_k|^{2}
-
\Re\{\delta_k^\ast\eta_k^{2}\}
}{
v_k^{2}-|\delta_k|^{2}
}
\right),
\end{equation}
which is valid when $v_k>|\delta_k|$.
Assuming conditional independence across frequencies, the negative log-likelihood ($\mathcal{L}_{\mathrm{NLL}}$) of the spectrum is thus
\begin{equation}
\mathcal{L}_{\mathrm{NLL}}
=
\sum_{k=1}^{K}
\left[
\frac{
v_k|\eta_k|^{2}
-
\Re\left\{\delta_k^\ast\eta_k^{2}\right\}
}{
v_k^{2}-|\delta_k|^{2}
}
+
\frac{1}{2}\ln\!\big(v_k^{2}-|\delta_k|^{2}\big)
+
\ln(\pi)
\right].
\label{eq:complex-nll-vector}
\end{equation}
When normalized uncertainty estimates $\Delta\tilde{\rho}_k'$ and $\Delta\tilde{\rho}_k''$ are available, the variance and pseudo-variance are
\begin{equation}
v_k = (\Delta\tilde{\rho}'_k)^{2} + (\Delta\tilde{\rho}''_k)^{2},
\qquad
\delta_k = (\Delta\tilde{\rho}'_k)^{2} - (\Delta\tilde{\rho}''_k)^{2}
+ 2i\,\operatorname{Cov}(\Delta\tilde{\rho}'_k,\Delta\tilde{\rho}''_k),
\end{equation}
respectively, where $\operatorname{Cov}(\Delta\tilde{\rho}'_k,\Delta\tilde{\rho}''_k)$ denotes the covariance between the real and imaginary uncertainty components. If uncertainties are unknown, $v_k$ and $\delta_k$ can be estimated from expectations over the residuals \citep[e.g.,][]{rybkin_simple_2021}. 

\subsubsection{Kullback--Leibler divergence}
The second term in Eq.~\eqref{eq:elbo} regularizes the latent space using a standard normal distribution prior. For each SIP spectrum, the $D_\mathrm{KL}$ between the approximate posterior and the prior is \citep{kingma_auto-encoding_2014}
\begin{equation}
D_{\mathrm{KL}}\left(
q_{\phi}\left(\mathbf{z}\,|\,\boldsymbol{\tilde{\rho}}\right)
\;\big\|\;
p(\mathbf{z})
\right)
=
\frac{1}{2}
\sum_{s=1}^{S}
\left(
\mu_{\phi,s}^{2}
+
\sigma_{\phi,s}^{2}
-
1
-
\ln \sigma_{\phi,s}^{2}
\right).
\label{eq:kld-single}
\end{equation}

\subsubsection{Training procedure}

The total loss function is the negative ELBO,
\begin{align}
\mathcal{L}(\theta,\phi;\boldsymbol{\tilde{\rho}},\mathbf{f})
&=
-\operatorname{ELBO}\left(\theta,\phi;\boldsymbol{\tilde{\rho}},\mathbf{f}\right),\\
&=
\mathcal{L}_{\mathrm{NLL}}
+
D_{\mathrm{KL}},
\label{eq:total_loss}
\end{align}
which is minimized by tuning the CVAE parameters with the PyTorch \texttt{AdamW} optimizer \citep{paszke_pytorch_2019}, using an initial learning rate of $10^{-3}$ and all other hyperparameters set to their default values. No additional weight is imposed on the $D_{\mathrm{KL}}$ term in Eq.~\eqref{eq:total_loss} and each iteration processes all 140 spectra in a single batch. Training is capped at $10^5$ iterations and stops when the mean loss improves by less than $1\,\%$ between consecutive windows of 1000 iterations.

\subsection{Error metrics}

Let $\mathbf{x}\in\mathbb{R}^{J\times K}$ denote a reference normalized SIP attribute, and let $\hat{\mathbf{x}}$ be its
prediction. We define the normalized mean absolute error (NMAE) as
\begin{equation}
\mathrm{NMAE}
=
100\,\left\langle \left|
\frac{\hat{\mathbf{x}}-\mathbf{x}}
{\max(\mathbf{x})-\min(\mathbf{x})}
\right| \right\rangle,
\label{eq:nmae}
\end{equation}
where the $\max$ and $\min$ operators are taken over samples and
frequencies, and where $\langle\cdot\rangle$ denotes averaging. Depending on the averaging domain, Eq.~\eqref{eq:nmae} yields a global, a frequency-dependent, or a sample-dependent error.

%%% DATA %%%
\section{Dataset} 
\label{sec:dataset}
The dataset consists of laboratory and field SIP measurements from unconsolidated and consolidated materials. All spectra are acquired using a SIP-Fuchs-III instrument at 19 logarithmically spaced frequencies between 90 mHz and 20 kHz. Each spectrum is measured twice under identical conditions, and the mean value and standard deviation are used as the measurement and uncertainty, respectively. Table~\ref{tab:dataset_summary} summarizes the data sources and sample counts. All 140 spectra are used to train the CVAE unsupervisedly, therefore the results assess performance only for these samples and do not establish transfer to unseen lithologies.

%TC:ignore
\begin{table}
\centering
\caption{Summary of the selected SIP datasets used in this study.}
\label{tab:dataset_summary}
\begin{tabular}{lllr}
\toprule
Data source & Grain or inclusion size & Reference & Number of samples \\
\midrule
Pyrite--sand mixtures & $\leq 0.5$, $0.5$--$1$, $1$--$3$ mm & This study & 30 \\
Graphite--sand mixtures & $\approx 44~\upmu\mathrm{m}$ & \citet{benzetta_caracterisation_2024} & 25 \\
Stainless steel spheres & $2$--$8$ mm & This study & 11 \\
Stainless steel cylinders & $2$--$6$ mm & This study & 6 \\
Canadian Malartic rock cores & Heterogeneous & \citet{berube_mineralogical_2019} & 12 \\
Highland Valley rock cores & Heterogeneous & \citet{grenon_caracterisation_2018} & 5 \\
Canadian Malartic field data & Heterogeneous & \citet{berube_mineralogical_2019} & 26 \\
Concrete samples & Heterogeneous & \citet{khajehnouri_validation_2020} & 25 \\
Total & & & 140 \\
\bottomrule
\end{tabular}
\end{table}
%TC:endignore

\subsection{Unconsolidated materials}

Polarizable grains embedded in clean feldspar sand are used to investigate the dependence of SIP on conductive phase properties and volumetric content. Samples are air-dried, poured without compaction, and saturated with KCl solutions prior to measurement. The first series consists of 30 mixtures with three pyrite size classes ($\leq0.5$ mm, 0.5--1~mm, 1--3~mm) and volume fractions from 1\,\% to 10\,\%. Measurements are performed in a cylindrical insulating cell (2.6~cm diameter, 7~cm length) using steel current electrodes and nonpolarizable potential electrodes. Samples are saturated with a KCl solution of 160~$\upmu$s~cm$^{-1}$ at 21~$^\circ$C, and the absence of sample holder polarization is verified through blank measurements. The second series includes 25 mixtures with graphite grains with a mean diameter of 44~$\upmu$m and volume fractions between 0.2\,\% and 5.0\,\% \citep{benzetta_caracterisation_2024}, prepared with a 37~$\upmu$s~cm$^{-1}$ KCl solution. We also include 17 measurements of stainless steel spheres and cylinders embedded in sand saturated with a 333~$\upmu$s~cm$^{-1}$ KCl solution. These inclusions have diameters ranging from 2 to 8~mm for spheres and 2 to 6~mm for cylinders. These measurements are performed in a sandbox using a miniature Schlumberger array with copper current electrodes spaced by 124~mm and Ag--AgCl potential electrodes spaced by 20~mm.

\subsection{Consolidated materials}

The dataset includes 17 previously published core sample spectra from the Highland Valley Cu and Canadian Malartic Au deposits, and 26 field SIP spectra from a north--south profile on the Bravo Zone at Canadian Malartic \citep{grenon_caracterisation_2018, berube_mineralogical_2019}. At Canadian Malartic, the core samples and field profile cover hydrothermally altered sedimentary rocks ranging from graywacke to mudstone, locally cut by felsic to intermediate intrusions and associated with sulfide mineralization \citep{berube_mineralogical_2019}. The Highland Valley samples originate from the Guichon Creek batholith, which hosts porphyry Cu mineralization \citep{grenon_caracterisation_2018}. The consolidated materials dataset is broadened with 25 spectra from alkali--silica reactive and nonreactive concrete samples cast simultaneously and cured for 24~h \citep{khajehnouri_measuring_2019, khajehnouri_validation_2020}. While the present dataset suffices to establish the methodology, the rock cores represent a relatively narrow lithological range, and future validation should include larger lithology-dependent datasets in which texture, formation factor, and petrophysical relationships can exert more diverse SIP controls \citep[e.g.,][]{johansson_spectral_2020,peshtani_permeability_2024,weller_prediction_2026}.

%%% MODEL SELECTION %%%
\section{Model selection}
\subsection{Latent space and decoder}
\begin{figure}
\centering
\subfloat[]{%
  \includegraphics[clip,width=0.49\textwidth]{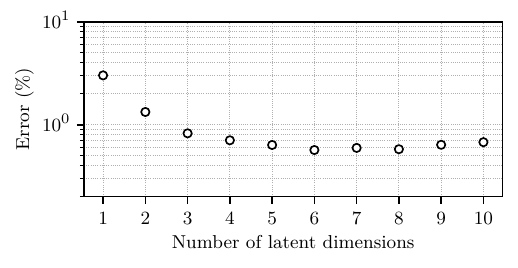}%
}
\vspace{0.5em}
\subfloat[]{%
  \includegraphics[clip,width=0.49\textwidth]{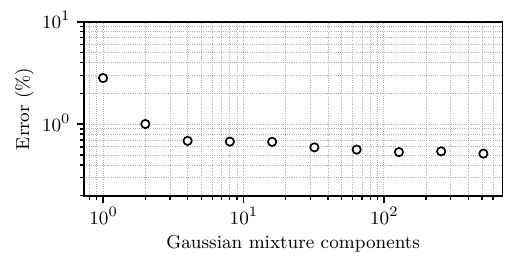}%
}
\caption{
Model selection results showing NMAE between measured and modelled complex resistivity as a function of (a) latent dimension $S$ and (b) number of Gaussian mixture components $N$. Values are averaged over real and imaginary components and three repetitions. Error bars are smaller than the markers.
}
\label{fig:model-selection}
\end{figure}

Figure~\ref{fig:model-selection} summarizes the NMAE between measured and modelled data using varying latent space dimensionality and number of Gaussian mixture components. To ensure that the latent space serves as an effective bottleneck, we first use a wide encoder with 32 nodes and a Gaussian mixture with 32 components. 

%TC:ignore
%TC:endignore

Figure~\ref{fig:model-selection} shows that reconstruction accuracy improves rapidly with both latent dimensionality and mixture size before reaching diminishing returns. Increasing $S$ from $1$ to $3$ reduces the NMAE on $\rho'$ from $2.85\,\%$ to $0.80\,\%$, with similar behaviour for $\rho''$. The lowest errors occur for $S=5$--$6$, where NMAE stabilizes at $0.60$--$0.63\,\%$ for $\rho'$ and $0.54$--$0.64\,\%$ for $\rho''$. Similarly, increasing the number of Gaussian mixture components from $N=1$ to $N=128$ reduces the NMAE from $2.56\,\%$ to $0.57\,\%$ for $\rho'$ and from $3.08\,\%$ to $0.50\,\%$ for $\rho''$. Using $N\geq256$ provide only minor gains. We therefore select $S=6$ and $N=128$.

\subsection{Encoder architecture}

\begin{figure}
\centering
\includegraphics[width=0.5\textwidth]{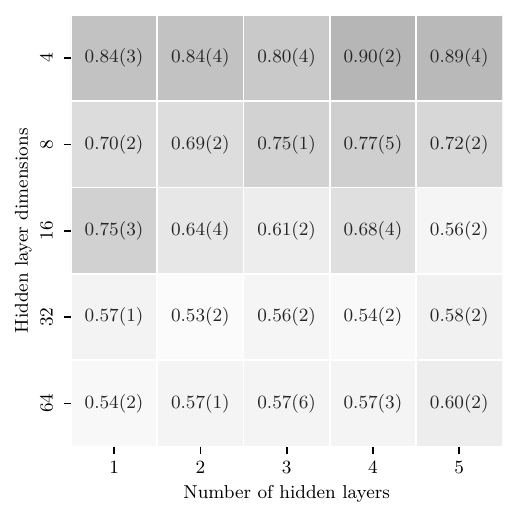}
\caption{
Reconstruction NMAE of complex resistivity spectra as a function of encoder depth and width. The values report the mean and standard deviation across three repetitions. The optimal architecture consists of two hidden layers with 32 units.
}
\label{fig:architecture}
\end{figure}

We perform a grid search by varying both the number of hidden layers and the width of each layer to identify the optimal encoder architecture. The results in Figure~\ref{fig:architecture} show that reconstruction error is primarily controlled by layer width. For example, with one hidden layer, the NMAE decreases from $0.84\,\%$ at dimension $4$ to $0.54\,\%$ at dimension $64$. Two-layer encoders follow the same trend and achieve the best overall performance at a width of $32$, yielding a reconstruction error of $0.53\,\%$. Increasing the width beyond $32$ does not improve accuracy for the two-layer case, nor does increasing the number of layers. We therefore select an encoder with two hidden layers of dimension $32$.

%TC:ignore
%TC:endignore

\subsection{Learning curves}
\begin{figure}
    \centering
    \includegraphics[width=0.5\textwidth]{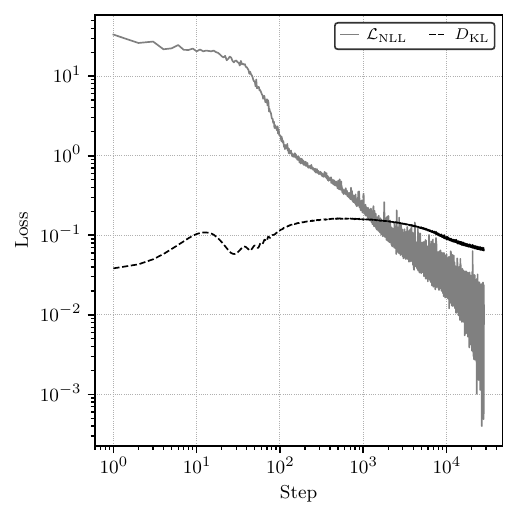}
\caption{
Learning curves of the selected CVAE showing the evolution of the negative log-likelihood ($\mathcal{L}_\mathrm{NLL}$) and Kullback--Leibler divergence ($D_\mathrm{KL}$) during training. The model converges to a stable solution, and training is stopped when the total loss no longer improves by more than 1\,\% over the last 1000 iterations.
}
    \label{fig:learning-curves}
\end{figure}

The evolution of the selected CVAE loss function during training is shown in Figure~\ref{fig:learning-curves} and is characteristic of stable optimization. During the first 500 steps, optimization prioritizes reconstruction, after which latent space regularization increases while the data misfit decreases more gradually. The $\mathcal{L}_\mathrm{NLL}$ term decreases from $3.33\times 10^{1}$ to $7.71\times 10^{-3}$ over $28\,272$ steps. The standard deviation over the last 1000 iterations is $3.84\times 10^{-3}$, which indicates convergence with small oscillations. The $D_\mathrm{KL}$ term evolves consistently with its regularization role. It varies between $3.84\times 10^{-2}$ and $6.75\times 10^{-2}$ and has a small standard deviation of $9.25\times 10^{-4}$ in the final 1000 iterations. 

%TC:ignore
%TC:endignore

%%% RESULTS %%%
\section{Results}
We evaluate the proposed method along seven criteria: (1) reconstruction accuracy, (2) learned RTD representation, (3) geophysical interpretability, (4) model sensitivity, (5) latent space organization, (6) generative capabilities, and (7) comparison with conventional methods. 

\subsection{Reconstruction accuracy}
\subsubsection{Representative spectrum reconstructions}
\begin{figure}
    \centering
    \includegraphics[width=0.5\textwidth]{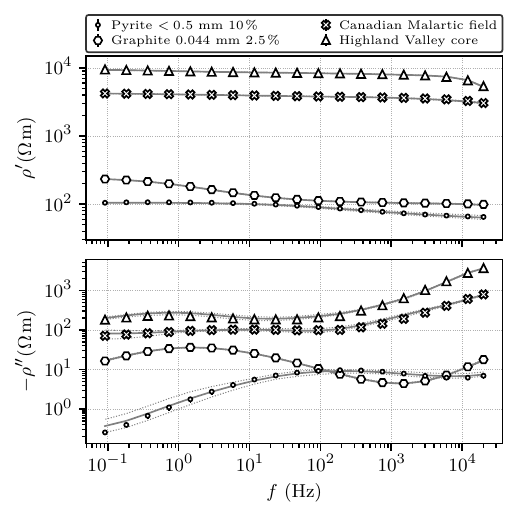}
\caption{
Examples of CVAE reconstructions for four SIP spectra drawn from laboratory and field measurements. Observed values are plotted as markers with vertical measurement error bars, which are smaller than the markers in most cases. The mean CVAE prediction is shown as a solid line and its $95\,\%$ posterior confidence interval as dotted lines. The examples span a wide range of SIP responses, including a pyrite--sand mixture, a graphite--sand mixture, and heterogeneous rock core and field measurements.
}
    \label{fig:fit}
\end{figure}

The examples in Figure~\ref{fig:fit} show that the CVAE reproduces a wide range of SIP responses with low reconstruction error. Across the selected samples, the NMAE for $\rho'$ remains between $0.46\,\%$ and $1.83\,\%$, and the NMAE for $\rho''$ lies between $0.22\,\%$ and $0.89\,\%$. In all cases, the predicted mean curves closely follow the observed data. The slightly larger NMAE for $\rho'$ is consistent with its larger measurement uncertainty relative to its range.

%TC:ignore
%TC:endignore

The dotted curves in Figure~\ref{fig:fit} correspond to the $95\,\%$ confidence interval of the CVAE. The influence of measurement uncertainty on model uncertainty is difficult to assess visually in Figure~\ref{fig:fit} because the vertical error bars, which represent the measurement standard deviations, are smaller than the markers at most frequencies. The response of model uncertainty to changes in the assumed measurement noise is therefore evaluated separately in Section~6.2.3.

\subsubsection{Error distribution across frequency}
\begin{figure}
    \centering
    \includegraphics[width=0.5\textwidth]{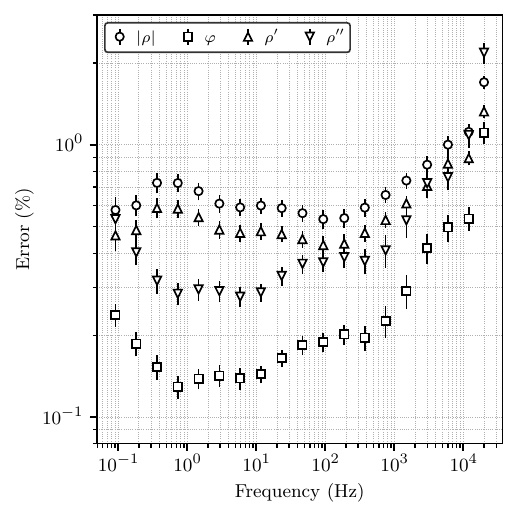}
\caption{
Frequency-dependent reconstruction error of the CVAE expressed as the NMAE. Errors are shown for the complex resistivity magnitude $|\rho|$, for its phase shift $\varphi$ and for its real ($\rho'$) and imaginary ($\rho''$) components. Error bars indicate the standard error of the mean across all samples and may be smaller than the markers at some frequencies.
}
\label{fig:error-analysis}
\end{figure}

Figure~\ref{fig:error-analysis} shows that the CVAE achieves reconstruction errors below 1\,\% across the 0.1~Hz to 1~kHz range, which contains the dominant interpretable polarization signatures. Errors are largest at the highest frequencies, reaching approximately $1.7\,\%$ for $|\rho|$ and $1.3$–$2.2\,\%$ for $\rho'$ and $\rho''$ at $20$~kHz, where capacitive coupling and instrumental effects typically limit SIP interpretability. The error for $\varphi$ also peaks in this range, with an NMAE close to $1.11\,\%$.

%TC:ignore
%TC:endignore

As frequency decreases, errors fall below $1\,\%$ and remain stable. In the low-frequency band between $0.1$~Hz and $1$~kHz, the NMAE ranges between $0.53$ and $0.73\,\%$ for $|\rho|$, $0.43$ and $0.59\,\%$ for $\rho'$, and $0.28$ and $0.41\,\%$ for $\rho''$. Over the same range, $\varphi$ is reconstructed with NMAE values between $0.13$ and $0.23\,\%$.

\subsubsection{Relaxation kernel comparison}
Considering this dataset, Debye decomposition with $c=1$ reduces the global reconstruction NMAE by $41$--$47\,\%$ across the four resistivity components relative to Warburg decomposition with $c=0.5$. Moreover, when $c$ is treated as a trainable decoder output constrained to $0<c<1$, the model converges to $c=1$ for this dataset.

\subsection{Learned representation of the RTD}

\subsubsection{RTD structure and variability}
\begin{figure}
    \centering
  \includegraphics[width=0.5\textwidth]{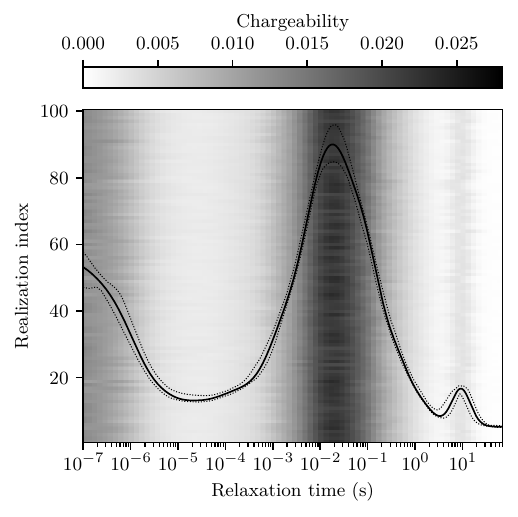}%
\caption{
Posterior RTD of the 2.5\,\% graphite--sand mixture. The carpet plot shows 100 posterior RTD realizations evaluated on the quadrature grid, where each row corresponds to a single realization and the grayscale intensity represents the dimensionless chargeability contribution of each quadrature bin. The black solid curve is the median continuous RTD from the Gaussian mixture model, and the dotted curves show its 95\,\% confidence interval. These curves are normalized to the height of the carpet plot for visualization purposes and their vertical scale is arbitrary.
}
    \label{fig:carpet}
\end{figure}

Figure~\ref{fig:carpet} shows the learned posterior distribution of the RTD for a representative SIP spectrum using both discrete and continuous representations. The carpet plot shows 100 posterior RTD realizations evaluated on the quadrature grid, with each row corresponding to one realization. The grayscale intensity represents the dimensionless chargeability contribution assigned to each quadrature bin. Summing the contributions yields the total chargeability $m_\theta$. The median continuous RTD and its 95\,\% confidence interval are superimposed on the carpet plot. 

%TC:ignore
%TC:endignore

The inferred RTD in Figure~\ref{fig:carpet} exhibits a stable and well-resolved multimodal structure. Across the posterior realizations, the total chargeability is $0.59 \pm 0.01$. A dominant mode is consistently recovered at $\tau^\ast \approx 1.8\times10^{-2}\,\mathrm{s}$, corresponding to a characteristic frequency of approximately $9$~Hz. This mode is interpreted as interfacial polarization of the conductive graphite grains in the electrolyte \citep{benzetta_caracterisation_2024}. The minor mode near $\tau^\ast \approx 9\,\mathrm{s}$ corresponds to a frequency below the measurement range. It may be attributed to the aggregation of some graphite grains into larger clumps in the sand mixture, which can produce a superposition of SIP responses \citep[e.g.,][]{martin_superposition_2023}. The response at $\tau^\ast < 10^{-7}\,\mathrm{s}$ is part of the instrument's high-frequency capacitive response, which is not fully described by the permittivity parameter.  Evaluating mode persistence across posterior realizations is a key advantage of the probabilistic formulation. For broad, weak, overlapping, or monotonic spectra, an RTD mode should be interpreted as physically meaningful only if it persists across realizations, lies within the resolved time range, and is supported by small reconstruction residuals. Conversely, broad intervals, unstable modes, or modes near the quadrature grid bounds indicate that the RTD should be interpreted cautiously.

\subsubsection{Dataset-level RTD organization}
\begin{figure}
    \centering
  \includegraphics[width=0.5\textwidth]{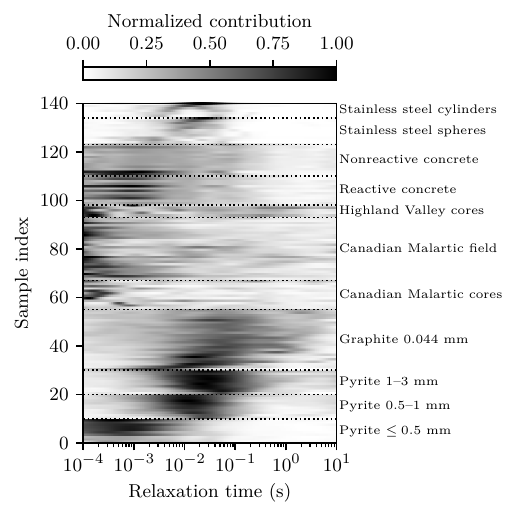}%
\caption{
Block-normalized heatmap of the posterior RTD across all samples. Each row corresponds to one spectrum, and grayscale indicates the contribution of Gaussian mixture components as a function of relaxation time. The normalization emphasizes structural RTD differences between sample groups, which are separated by horizontal dotted lines. Grain and inclusion sizes are reported in Table~\ref{tab:dataset_summary}.
}
    \label{fig:rtd-heatmap}
\end{figure}

Analysis of the RTDs across all 140 samples reveals a consistent valley at $\tau=100~\upmu$s separating the high- and low-frequency responses. We focus on the low-frequency response, which corresponds to relaxation times between $100~\upmu$s and $10$~s. Within this range, distinct RTD signatures emerge for each material class, with within-series variability remaining lower than between-series variability, as evidenced in Figure~\ref{fig:rtd-heatmap}. 

%TC:ignore
%TC:endignore

The persistence of the RTD peak across posterior realizations, as exemplified in Figure~\ref{fig:carpet}, and the recurring patterns within material groups in Figure~\ref{fig:rtd-heatmap} support the consistency of the inferred SIP parameters for the studied dataset. However, these observations do not uniquely identify the mechanisms responsible for the RTD peaks. Relaxation times can be compared between minerals, but their differences cannot be attributed to grain size alone because they also depend on mineral electrical properties, particle shape, and interstitial water conductivity \citep[e.g.,][]{misra_interfacial_2016, wu_effects_2020}.

\subsubsection{Sensitivity to measurement uncertainty}
To test whether the inferred uncertainty responds to the assumed measurement quality, we train three otherwise identical CVAEs using the recorded measurement standard deviations, $1.5$ times these values, and twice these values. For each model, we generate 100 realizations per spectrum and average the resulting posterior standard deviations across all 140 spectra and frequencies. Relative to the baseline model, the average posterior standard deviation of $\tilde{\rho}'$ increases by $5\,\%$ and $11\,\%$ at uncertainty scales of $1.5$ and $2$, respectively. The corresponding standard deviation of $m_{[10^{-4},10^1]}$ increases by approximately $10\,\%$ at both scales. In comparison, the changes for $\tilde{\rho}''$ are $6\,\%$ and $3\,\%$, while the variability of $\bar{\tau}_{[10^{-4},10^1]}$ does not change monotonically. The assumed measurement uncertainty therefore affects the posterior distribution, but the response is modest and depends on the predicted quantity.

\subsection{Interpretability of the learned representation}
\subsubsection{Integrating parameters}
\begin{figure}
    \centering
    \includegraphics[width=0.5\textwidth]{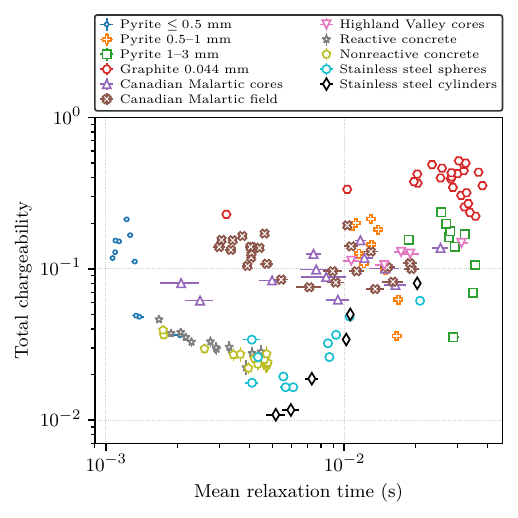}
\caption{Total chargeability $m_{[10^{-4},10^1]}$ as a function of mean relaxation time $\bar{\tau}_{[10^{-4},10^1]}$ for all geomaterial groups in the dataset. Each marker corresponds to a single SIP spectrum, colored by sample group. The distribution of points highlights systematic contrasts between pyrite-- and graphite--sand mixtures, natural rock samples, concrete, and stainless steel inclusions in terms of both polarization strength and characteristic timescale.}
\label{fig:tau_vs_m}
\end{figure}

The joint distribution of truncated mean relaxation time $\bar{\tau}_{[10^{-4},10^1]}$ and total chargeability $m_{[10^{-4},10^1]}$ across all material groups reveals a weak separation between material classes, as evidenced in Figure~\ref{fig:tau_vs_m}. 

%TC:ignore
%TC:endignore

Across all pyrite grain sizes, the average $m_{[10^{-4},10^1]}$ remains close to $\approx 0.13$, whereas the average $\bar{\tau}_{[10^{-4},10^1]}$ increases from $1.3\times10^{-3}$~s for the finest grains to $2.9\times10^{-2}$~s for the coarsest ones. Graphite mixtures occupy a distinct region of the parameter space in Figure~\ref{fig:tau_vs_m}, with both higher chargeability and longer relaxation times than any of the pyrite mixtures. The average $m_{[10^{-4},10^1]}$ for graphite is $\approx 0.38$, with an average $\bar{\tau}_{[10^{-4},10^1]}$ on the order of $10^{-2}$~s. Although the same nominal grain size is used throughout the graphite mixtures, their volumetric fraction varies from $0.2$ to $5.0\,\%$, which may change the connectivity of the graphite grains and therefore contribute to the variability of their SIP response \citep[e.g.,]{abdulsamad_complex_2019, benzetta_caracterisation_2024}. 

Table~\ref{tab:tau-m-summary} summarizes the group-level means and standard deviations of the estimated RTD parameters. The similar mean chargeabilities of the three pyrite series are consistent with their common volumetric-fraction range of $1$--$10\,\%$. For disseminated electronic conductors, chargeability is controlled primarily by volumetric fraction, whereas grain size mainly controls relaxation time \citep{gurin_time_2013}. Larger pyrite grains are therefore not expected, by themselves, to produce higher chargeability. The higher $m_{[10^{-4},10^1]}$ of the graphite mixtures indicates a stronger polarization response under the experimental conditions used for that series.

%TC:ignore
\begin{table}
\centering
\caption{
Group-level means and standard deviations (s.d.) of the truncated mean relaxation time $\bar{\tau}_{[10^{-4},10^1]}$ and total chargeability $m_{[10^{-4},10^1]}$.
}
\label{tab:tau-m-summary}
\begin{tabular}{lcccc}
\toprule
Sample group & $\bar{\tau}$ mean (s) & $\bar{\tau}$ s.d. (s) & $m$ mean & $m$ s.d. \\
\midrule
Pyrite $\leq 0.5$ mm & $1.31\times 10^{-3}$ & $2.89\times 10^{-4}$ & 0.117 & 0.058 \\
Pyrite 0.5--1 mm & $1.35\times 10^{-2}$ & $2.15\times 10^{-3}$ & 0.137 & 0.061 \\
Pyrite 1--3 mm & $2.89\times 10^{-2}$ & $4.74\times 10^{-3}$ & 0.145 & 0.060 \\
Graphite 0.044 mm & $2.76\times 10^{-2}$ & $8.03\times 10^{-3}$ & 0.377 & 0.088 \\
Canadian Malartic cores & $1.02\times 10^{-2}$ & $6.40\times 10^{-3}$ & 0.099 & 0.029 \\
Canadian Malartic field & $8.17\times 10^{-3}$ & $5.31\times 10^{-3}$ & 0.122 & 0.032 \\
Highland Valley cores & $1.85\times 10^{-2}$ & $7.61\times 10^{-3}$ & 0.124 & 0.016 \\
Reactive concrete & $2.97\times 10^{-3}$ & $1.02\times 10^{-3}$ & 0.033 & 0.006 \\
Nonreactive concrete  & $3.79\times 10^{-3}$ & $1.15\times 10^{-3}$ & 0.027 & 0.005 \\
Stainless steel spheres & $7.99\times 10^{-3}$ & $4.74\times 10^{-3}$ & 0.030 & 0.014 \\
Stainless steel cylinders & $9.89\times 10^{-3}$ & $5.42\times 10^{-3}$ & 0.034 & 0.027 \\
\bottomrule
\end{tabular}
\end{table}
%TC:endignore

Natural rock samples from the Canadian Malartic and Highland Valley deposits cluster around $10^{-2}$~s, with $m_{[10^{-4},10^1]}$ values between $0.10$ and $0.12$. Core and field measurements from the Canadian Malartic deposit overlap in Figure~\ref{fig:tau_vs_m}, despite their different acquisition conditions. The concrete samples are characterized by short relaxation times and low chargeability. Reactive and nonreactive concretes both exhibit $\bar{\tau}$ values of a few $10^{-3}$~s and $m < 0.04$, with reactive concrete showing slightly higher chargeability. Lastly, stainless steel spheres with diameters of $2$--$8$~mm and cylinders with diameters of $2$--$6$~mm have average $\bar{\tau}_{[10^{-4},10^1]}$ values of $8.0\times10^{-3}$ and $9.9\times10^{-3}$~s, respectively, which are closest to that of the $0.5$--$1$~mm pyrite series. These values do not imply equivalent particle sizes because relaxation time also depends on material properties and geometry.

\subsubsection{Total chargeability}
\begin{figure}
\centering
\includegraphics[width=0.5\textwidth]{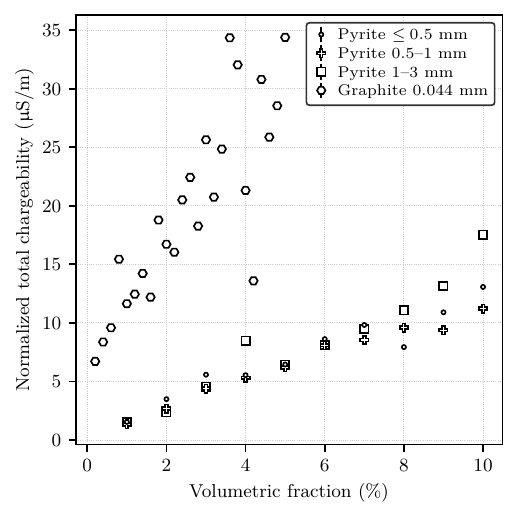}
\caption{
Truncated normalized total chargeability $m_{[10^{-4},10^1]}/\rho_0$ as a function of volumetric fraction for pyrite and graphite sand mixtures. Markers denote different grain size series, and error bars represent posterior standard deviations, which are generally smaller than the markers. The data reveal an approximately linear increase in normalized chargeability with polarizable mineral content, with a markedly higher sensitivity to graphite than to pyrite.
}
\label{fig:chargeability}
\end{figure}

Figure~\ref{fig:chargeability} shows the relationship between truncated total chargeability normalized by the direct current resistivity, $m_{[10^{-4},10^1]}/\rho_0$, and volumetric fraction for the pyrite–sand and graphite–sand mixtures. For these materials, the normalized chargeability increases linearly with volumetric fraction. The strength of this relationship is high across all pyrite mixtures, with coefficients of determination ranging from 0.94 to 0.98.

%TC:ignore
%TC:endignore

The three pyrite series exhibit similar slopes and intercepts that remain close to zero. The $\leq 0.5$~mm pyrite grains yield a slope of $1.15 \pm 0.10~\upmu\mathrm{S/m}$ per~\%, whereas the 0.5--1~mm grains yield a slope of $1.15 \pm 0.05~\upmu\mathrm{S/m}$ per~\%. The 1--3~mm grains produce a slope of $1.47 \pm 0.13~\upmu\mathrm{S/m}$ per~\%.  Contrastingly, the graphite–sand mixtures follow a linear trend with a slope of $3.22 \pm 0.75~\upmu\mathrm{S/m}$ per~\%. The difference between the graphite and pyrite slopes may arise from the nature of the polarizable mineral, the different electrolyte conductivities used for the two series, or both. Moreover, graphite grain connectivity in the matrix affects chargeability \citep{abdulsamad_complex_2019, benzetta_caracterisation_2024}, which may explain the variability of the graphite responses and the lower determination coefficient of 0.45.

\subsubsection{Mean relaxation time}
\begin{figure}
\centering
\includegraphics[width=0.5\textwidth]{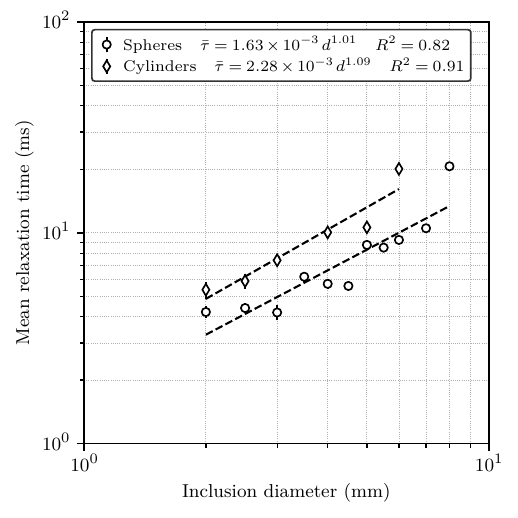}
\caption{
Truncated mean relaxation time $\bar{\tau}_{[10^{-4},10^1]}$, in ms, as a function of inclusion diameter ($d$), in mm, for stainless steel spheres and cylinders in feldspar sand. Error bars represent posterior standard deviations and are smaller than the markers for most samples. The data show a systematic increase of $\bar{\tau}$ with inclusion size for both geometries.
}
\label{fig:tau}
\end{figure}

Figure~\ref{fig:tau} illustrates the relationship between $\bar{\tau}_{[10^{-4},10^1]}$ and inclusion diameter for stainless steel spheres and cylinders. For both geometries, relaxation time increases with inclusion size, which indicates that the polarization timescale recovered by the CVAE is controlled by the dimension of the inclusions. The data are described by a power-law relationship of the form $\bar{\tau} = \beta\,d^{\alpha}$, as evidenced by strong correlations in log--log space for both geometries.

%TC:ignore
%TC:endignore

For spherical inclusions, the fitted scaling exponent is $\alpha = 1.01 \pm 0.16$, with a coefficient of determination of $0.82$ and a root-mean-square error of $2.36\times10^{-3}$~s. The exponent indicates an approximately linear scaling between the mean relaxation time and sphere diameter over the investigated range from 2 to 8~mm. The corresponding prefactor is $\beta = 1.63\times10^{-3}$~s\,mm$^{-\alpha}$. Cylindrical inclusions show a similar dependence, with a fitted exponent of $\alpha = 1.09 \pm 0.18$ and a higher determination coefficient of $0.91$. The prefactor for cylinders, $\beta = 2.28\times10^{-3}$~s\,mm$^{-\alpha}$, is of the same order of magnitude as that obtained for spheres. The observed linear scaling agrees with the mechanistic model of \cite{misra_interfacial_2016}, which predicts Debye-type relaxation.

\subsection{Sensitivity analysis}

\begin{figure}
    \centering
    \includegraphics[width=0.5\textwidth]{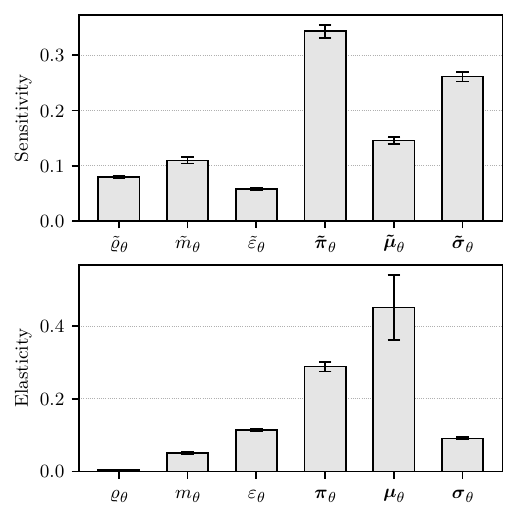}
\caption{
Relative sensitivity structure of the CVAE parameters. The top panel shows sensitivity indices from the Jacobian with respect to the raw RTD parameters. The bottom panel shows elasticities with respect to the logarithm of the physical RTD parameters. Error bars denote the standard error across samples.
}
\label{fig:sensitivity}
\end{figure}

We analyze the sensitivity of the CVAE using the Jacobian of the end-to-end mapping from SIP data to RTD parameters \citep[see][]{berube_bayesian_2023}. The sensitivities with respect to the raw decoder outputs and the elasticities with respect to the logarithm of the physical RTD parameters are reported in Figure~\ref{fig:sensitivity}. In both cases, indices are normalized to sum to $100\,\%$. The raw sensitivities show that the decoder is primarily controlled by the mixture weights $\boldsymbol{\tilde{\pi}}_\theta$ (34.3\,\%), widths $\boldsymbol{\tilde{\sigma}}_\theta$ (26.1\,\%), and locations $\boldsymbol{\tilde{\mu}}_\theta$ (14.6\,\%), which together account for about 75\,\% of the total sensitivity. The influence of total chargeability $\tilde{m}_\theta$ is moderate (11.0\,\%), whereas those of the resistivity $\tilde{\varrho}_\theta$ and high-frequency $\tilde{\varepsilon}_\theta$ parameters are smaller, at 8.1\,\% and 5.9\,\%, respectively.

%TC:ignore
%TC:endignore

The dominant elasticity contributions arise from the Gaussian mixture locations $\boldsymbol{\mu}_\theta$ (45.1\,\%) and weights $\boldsymbol{\pi}_\theta$ (28.8\,\%), followed by the high-frequency parameter $\varepsilon_\theta$ (11.5\,\%), mixture widths $\boldsymbol{\sigma}_\theta$ (9.1\,\%), and total chargeability $m_\theta$ (5.1\,\%). The contribution of $\varrho_\theta$ is negligible (0.4\,\%). The difference between the raw and physical rankings results from the nonlinear transformations used to enforce parameter bounds and positivity. The dominance of location and weight parameters reflects their role in controlling how chargeability is distributed among the relaxation times, whereas $m_\theta$ applies a single global scaling to the RTD. The lower relative sensitivity of $m_\theta$ therefore does not mean that chargeability is not important or poorly recovered, but that variations among the SIP data are represented primarily through changes in RTD shape.

\subsection{Latent space representation}
\begin{figure}
\centering
\includegraphics[width=0.5\textwidth]{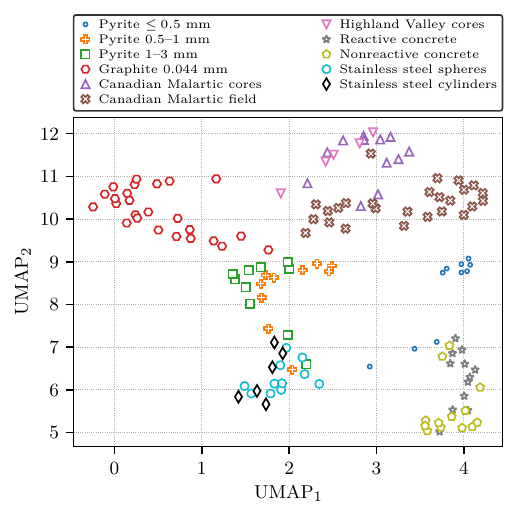}
\caption{
Two-dimensional uniform manifold approximation and projection (UMAP) of the CVAE latent space. Each marker represents one SIP spectrum, colored by sample group. The embedding shows compact and well-separated clusters despite the absence of class labels during training.
}
\label{fig:umap}
\end{figure}

We analyze the learned latent representation using a two-dimensional uniform manifold approximation and projection (UMAP; \citealt{mcinnes_umap:_2018}). The embedding, shown in Figure~\ref{fig:umap}, is computed without using class labels and therefore reflects similarities learned unsupervisedly from the data. The latent variables are not physical parameters but abstract features that represent variability in the SIP spectra. Nevertheless, the separation of distinct material groups into clusters indicates that the latent variables capture differences in the shape of the SIP spectra.

%TC:ignore
%TC:endignore

The pyrite–sand mixtures occupy neighbouring but distinct regions of the embedding (Figure~\ref{fig:umap}). As grain size increases, the cluster centroids shift systematically, indicating that the latent representation captures variations in SIP response. Graphite mixtures form a broader cluster despite their common nominal grain size. This is explained by the fact that graphite fraction varies from $0.2$ to $5.0\,\%$, which changes particle spacing and connectivity and thereby produces variability in the SIP responses of the mixtures. Natural rock samples form coherent groups, with Canadian Malartic core and field measurements occupying adjacent but distinct regions. This separation may reflect differences in acquisition conditions. Highland Valley core samples cluster near those of Canadian Malartic. Lastly, concrete samples form a separate cluster with a small spread, and reactive and nonreactive concretes are distinguishable.

\subsection{Generative properties}
\begin{figure}
    \centering
    \includegraphics[width=0.5\textwidth]{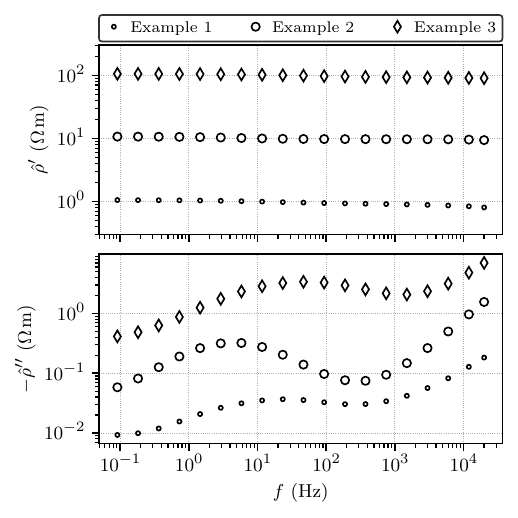}
\caption{
Examples of synthetic complex resistivity spectra generated by sampling the CVAE latent space. Each realization corresponds to a randomly drawn latent vector and is scaled by a direct current resistivity $\rho_0$. The top panel shows $\hat{\rho}'$ and the bottom panel shows $-\hat{\rho}''$ as functions of frequency $f$.
}
\label{fig:generative}
\end{figure}

The CVAE defines a generative model that produces synthetic SIP data by sampling latent variables $\mathbf{z}\sim\mathcal{N}(\mathbf{0},\mathbf{I}_{S})$ and decoding them into complex resistivity responses. Figure~\ref{fig:generative} shows three realization examples. The generated spectra reproduce key features of the training data, including the monotonic frequency dependence of $\rho'$ and broad polarization peaks in $-\rho''$, while their amplitudes and characteristic frequencies vary among realizations. The increase in $-\rho''$ at high frequencies arises from the permittivity term included in the forward model. These examples show that the CVAE generates plausible spectra based on the training data, although individual realizations do not represent specific polarization mechanisms.

%TC:ignore
%TC:endignore

\subsection{Comparison with conventional methods}

We compare the proposed approach with conventional deterministic and stochastic decomposition methods. The CVAE introduces a fundamentally different formulation that enables dataset-level decomposition, uncertainty quantification, and representation learning. The purpose of this comparison is to assess whether these new capabilities are achieved without compromising, and potentially improving, metrics such as reconstruction accuracy, parameter estimation, clustering performance, and computational cost. For comparisons with a deterministic method, we reproduce the approach of \cite{nordsiek_new_2008} while adding the constraint that chargeability does not exceed one. For comparisons with a stochastic approach, we use the code of \citet{berube_bayesian_2017} with a fifth-degree polynomial RTD.

\subsubsection{Reconstruction error}
Table~\ref{tab:dd_comparison} reports the NMAE for the real, imaginary, magnitude, and phase components of the complex resistivity. All methods are evaluated on the same dataset of 140 spectra. The CVAE yields mean NMAEs below $1\,\%$ for all four components. Compared with the deterministic decomposition, its mean NMAE is higher by approximately $0.07\,\%$ for both $\rho'$ and $|\rho|$, although the corresponding standard deviations across samples are approximately half as large. Conversely, the CVAE reduces the mean NMAE for $\rho''$ and $\varphi$. The reduction is supported more strongly for $\rho''$ ($p$-value $=7\times10^{-3}$) than for $\varphi$ ($p$-value $=6\times10^{-2}$). The CVAE shows more consistent performance across input spectra than the stochastic approach, as indicated by the smaller standard deviations in Table~\ref{tab:dd_comparison}. These variations reflect the different optimization objectives. In fact, both conventional methods fit each spectrum independently by separating their real and imaginary parts, whereas the CVAE jointly learns a probabilistic representation of the complete dataset according to both reconstruction and latent space regularization objectives. 

\begin{table}
\centering
\caption{
Comparison of reconstruction error between the proposed CVAE and conventional deterministic \citep{nordsiek_new_2008} and stochastic \citep{berube_bayesian_2017} decomposition methods, expressed as NMAE for the real ($\rho'$), imaginary ($\rho''$), magnitude ($|\rho|$), and phase ($\varphi$) components. Values are reported as mean $\pm$ standard deviation across the 140 samples.
}
\begin{tabular}{lrrrr}
\toprule
 & $\rho'$ (\%) & $\rho''$ (\%) & $|\rho|$ (\%) & $\varphi$ (\%) \\
\midrule
\cite{nordsiek_new_2008} & $0.52\pm0.64$ & $0.65\pm0.99$ & $0.66\pm0.83$ & $0.32\pm0.45$ \\
\cite{berube_bayesian_2017} & $0.57\pm0.99$ & $2.33\pm5.11$ & $0.97\pm2.15$ & $1.47\pm3.80$ \\
CVAE (this work) & $0.59\pm0.29$ & $0.53\pm0.37$ & $0.73\pm0.35$ & $0.27\pm0.20$ \\
\bottomrule
\end{tabular}
\label{tab:dd_comparison}
\end{table}

\subsubsection{Recovered parameters}

The coefficients of determination quantifying the relationship between pyrite content and $m_{[10^{-4},10^1]}$ for each decomposition method are reported in Table~\ref{tab:r2_pyrite_chargeability}. The CVAE enhances the recovery of relationships between chargeability and pyrite content across all grain size classes, with determination coefficient gains of $0.021$, $0.010$, and $0.012$ for the $\leq 0.5$~mm, $0.5$--$1$~mm, and $1$--$3$~mm classes, respectively. 

\begin{table}
\centering
\caption{
Comparison of the coefficient of determination between pyrite content and total chargeability for the conventional approach of \cite{nordsiek_new_2008} and the proposed CVAE method. Results are reported for different particle size classes.
}
\begin{tabular}{lrr}
\toprule
 & \cite{nordsiek_new_2008} & CVAE (this work) \\
\midrule
Pyrite $\leq 0.5$ mm & 0.923 & 0.944 \\
Pyrite 0.5--1 mm     & 0.973 & 0.983 \\
Pyrite 1--3 mm       & 0.927 & 0.939 \\
\bottomrule
\end{tabular}
\label{tab:r2_pyrite_chargeability}
\end{table}

\subsubsection{Clustering performance}
Table~\ref{tab:clustering_metrics} compares the separation of the 11 sample groups in the conventional RTD parameter space and in the UMAP projection of the CVAE latent space. The silhouette score measures within-group cohesion relative to between-group separation, whereas the Calinski--Harabasz index compares between-group and within-group dispersion. For both metrics, higher values indicate better clustering. The increase in silhouette score from 0.075 to 0.182 and in Calinski--Harabasz index from 58 to 132 therefore indicates that the latent representation separates the material groups more clearly than $m$ and $\bar{\tau}$ alone. However, the relatively low silhouette score also indicates that some groups remain overlapping. Sample labels are used only to calculate these evaluation metrics and are not used by the CVAE or UMAP.

%TC:ignore
\begin{table}
\centering
\caption{
Clustering metrics for conventional RTD parameters and the latent space projection.
}
\begin{tabular}{lrrr}
\toprule
Feature space & Silhouette & Calinski–Harabasz \\
\midrule
Debye parameters & 0.075  & 58 \\
UMAP latent projection & 0.182 & 132 \\
\bottomrule
\end{tabular}
\label{tab:clustering_metrics}
\end{table}
%TC:endignore

\subsubsection{Computation times}

All three methods are computationally inexpensive and readily executed on a personal computer. Sequential application of the deterministic decomposition to the 140 spectra requires $362 \pm 7$~s, while the stochastic method requires $295.4 \pm 0.9$~s using 64 walkers and 1000 iterations. In comparison, CVAE inference for the full dataset takes only $3.0 \pm 0.3$~ms, but training the CVAE takes approximately 271~s and must be performed once prior to inference. All timings are obtained on an Apple M5 chip and all methods run on CPU. 

\subsection{Practical implications and limitations}
The CVAE can serve as a survey-scale decomposition method for field SIP data. Once trained on a dataset using a common frequency grid, it infers the RTD and a latent representation of each spectrum. The resulting parameters can then be compared across measurement locations and survey times. For instance, SIP pseudosections or inversion results may be clustered based on their latent representations. In time-lapse monitoring applications, the trajectory of a measurement away from its baseline cluster may reveal a progressive change in its SIP signature. In both field and laboratory contexts, spectra lying far from the principal latent space clusters may also be considered potential anomalies. Similar uses of latent representations for monitoring and anomaly detection have been demonstrated in other fields \citep{costa_variational_2022, nicolaou_identifying_2026}. However, one cannot identify the specific polarization processes responsible for these patterns using the latent representation alone. Such an interpretation requires supporting geological and petrophysical information.

The CVAE is trained and evaluated on a common frequency grid, and the effect of varying acquisition protocols has not been evaluated. Spectra acquired using different frequency sampling can be interpolated onto a common grid over the frequency range covered by both datasets. However, we do not recommend extrapolating SIP data outside of the measured frequency range. A further limitation is that the CVAE is difficult to benchmark against conventional probabilistic decomposition. The stochastic approach of \citet{berube_bayesian_2017} samples the RTD posterior independently for each spectrum, whereas the CVAE approximates the posterior through a latent representation learned jointly from the complete dataset. Although the trained CVAE enables rapid inference over large numbers of spectra, its training cost must be included in computational comparisons. More importantly, the reconstruction errors and computation times reported here do not assess the calibration or coverage of the resulting uncertainty intervals. Such a comparison is difficult for the present dataset because the ground truth RTD of the investigated materials are unknown.

\section{Conclusions}
We formulate SIP decomposition as a learned inverse problem and show that a CVAE provides an effective tool for estimating the RTD of geomaterials. The results demonstrate that the model learns stable RTD representations, achieves data reconstruction errors below 1\,\%, recovers interpretable parameters, and automatically organizes the data in its latent space. Unlike conventional methods that impose discretization, smoothness, or parametric constraints, the CVAE represents the RTD as a continuous function learned from SIP data and provides a distribution of admissible solutions through its probabilistic decoder. By formulating SIP decomposition as an inverse mapping at the dataset level, the method supports clustering, discrimination, and exploratory characterization of SIP measurements, while also enabling potential applications in monitoring and anomaly detection. Moreover, the CVAE formulation preserves the coupling between the real and imaginary parts of SIP data, whereas conventional decomposition implementations treat these components independently. Results show that the CVAE yields reductions in reconstruction error for the imaginary component and phase, while also improving parameter recovery relative to conventional SIP decomposition methods. However, the learned latent variables remain abstract and do not correspond directly to physical parameters. In addition, curve fitting performance depends on the training data uncertainty, and extrapolation of this unsupervised model to unseen data may yield incorrect RTDs. Future work should examine geological conditioning, supervised training on larger datasets, generalization to unseen data, and applications in monitoring and anomaly detection.

\section*{Author statement}
\textbf{Charles Bérubé:} Conceptualization, Methodology, Software, Validation, Formal analysis, Investigation, Resources, Data curation, Writing - Original Draft, Writing - Review \& Editing, Visualization, Supervision, Project administration, Funding acquisition. 
\textbf{Sébastien Gagnon:} Investigation, Data curation, Software, Writing - Original Draft, Writing - Review \& Editing. 
\textbf{Lahiru Nagasingha:} Software, Writing - Review \& Editing. 
\textbf{Jean-Luc Gagnon:} Writing - Review \& Editing.
\textbf{Rachel Kenko:} Writing - Review \& Editing.
\textbf{Reza Ghanati:} Writing - Review \& Editing. 
\textbf{Frédérique Baron:} Conceptualization, Visualization, Writing – Review \& Editing.

\section*{Declaration of interests}
The authors declare that they have no known competing financial interests or personal relationships that could have appeared to influence the work reported in this paper.

\section*{Data and code availability}
The code is developed and the results are produced using Python 3.13.2 and PyTorch 2.7.0. Data processing, analysis, and visualization use NumPy 2.2.4, SciPy 1.15.2, pandas 2.2.3, Matplotlib 3.10.0, scikit-learn 1.6.1, seaborn 0.13.2, and umap-learn 0.5.9.post2. The dataset, trained model weights, and source code are available at \url{https://github.com/clberube/sip-decomposition-net}.

%%% ACKNOWLEDGMENTS %%%
\section*{Acknowledgments}
C.~L.~Bérubé acknowledges FRQ-NT Grant No. 326054 and NSERC Grant No. 2024-06161. S.~Gagnon is supported by an NSERC Undergraduate Student Research Award. The authors thank the editors, André Revil and the anonymous reviewers for constructive comments that helped improve the manuscript.

%TC:ignore
\appendix

\section{Likelihood function}
\label{app:complex-gaussian}

Let $\eta\in\mathbb{C}$, the SIP data residuals, denote a zero-mean complex random variable. We omit the frequency indices for clarity. Its second-order
statistics are characterized by the variance
\begin{equation}
v = \mathbb{E}[|\eta|^{2}],
\end{equation}
and the pseudo-variance
\begin{equation}
\delta = \mathbb{E}[\eta^{2}].
\end{equation}
Following \citet{picinbono_second-order_1996}, we define the augmented random vector
\begin{equation}
\underline{\eta}
=
\begin{bmatrix}
\eta\\
\eta^{*}
\end{bmatrix},
\end{equation}
whose augmented covariance matrix reads
\begin{equation}
\underline{\mathbf{C}}
=
\mathbb{E}\!\left[
\underline{\eta}\,\underline{\eta}^{\mathrm{H}}
\right]
=
\begin{bmatrix}
v & \delta\\
\delta^{*} & v
\end{bmatrix}.
\end{equation}
Positive definiteness of $\underline{\mathbf{C}}$ requires $v>|\delta|$. The probability density function of $\eta$ is then given by
\begin{equation}
p(\eta)
=
\frac{1}{\pi\sqrt{\det(\underline{\mathbf{C}})}}
\exp\!\left(
-\frac{1}{2}\underline{\eta}^{\mathrm{H}}
\underline{\mathbf{C}}^{-1}
\underline{\eta}
\right).
\label{eq:augmented-density}
\end{equation}
For the scalar case, the determinant and inverse of $\underline{\mathbf{C}}$ are
\begin{equation}
\det(\underline{\mathbf{C}})
=
v^{2}-|\delta|^{2},
\end{equation}
and
\begin{equation}
\underline{\mathbf{C}}^{-1}
=
\frac{1}{v^{2}-|\delta|^{2}}
\begin{bmatrix}
v & -\delta\\
-\delta^{*} & v
\end{bmatrix}.
\end{equation}
The quadratic form in the exponent of Eq.~\eqref{eq:augmented-density} is thus
\begin{align}
\underline{\eta}^{\mathrm{H}}
\underline{\mathbf{C}}^{-1}
\underline{\eta}
&=
\frac{2}{v^{2}-|\delta|^{2}}
\left(
v|\eta|^{2}
-
\Re\!\left\{\delta^{*}\eta^{2}\right\}
\right).
\end{align}
Substituting these results into Eq.~\eqref{eq:augmented-density} yields the scalar density
\begin{equation}
p(\eta;v,\delta)
=
\frac{1}{\pi\sqrt{\,v^{2}-|\delta|^{2}\,}}
\exp\!\left(
-\frac{
v|\eta|^{2}
-
\Re\!\left\{\delta^\ast\eta^{2}\right\}
}{
v^{2}-|\delta|^{2}
}
\right).
\end{equation}
%TC:endignore

\bibliographystyle{cas-model2-names} 
%TC:ignore
\bibliography{references}

@inproceedings{kingma_auto-encoding_2014,
	title = {Auto-{Encoding} {Variational} {Bayes}},
	booktitle = {2nd {International} {Conference} on {Learning} {Representations}, {ICLR} 2014, {Banff}, {AB}, {Canada}, {April} 14-16, 2014, {Conference} {Track} {Proceedings}},
	author = {Kingma, Diederik P. and Welling, Max},
	editor = {Bengio, Yoshua and LeCun, Yann},
	year = {2014},
	pages = {1--14},
}

@article{wu_effects_2020,
	title = {Effects of salinity and {pH} on the spectral induced polarization signals of graphite particles},
	volume = {221},
	issn = {0956-540X},
	doi = {10.1093/gji/ggaa087},
	number = {3},
	urldate = {2021-03-11},
	journal = {Geophysical Journal International},
	author = {Wu, Yuxin and Peruzzo, Luca},
	month = jun,
	year = {2020},
	pages = {1532--1541},
}

@article{martin_superposition_2023,
	title = {Superposition of induced polarization signals measured on pyrite–sand mixtures},
	volume = {234},
	issn = {0956-540X},
	url = {https://doi.org/10.1093/gji/ggad091},
	doi = {10.1093/gji/ggad091},
	number = {1},
	urldate = {2026-10-06},
	journal = {Geophysical Journal International},
	author = {Martin, Tina and Weller, Andreas},
	month = jul,
	year = {2023},
	pages = {699--711},
}

@article{costa_variational_2022,
	title = {Variational encoding approach for interpretable assessment of remaining useful life estimation},
	volume = {222},
	issn = {0951-8320},
	doi = {10.1016/j.ress.2022.108353},
	urldate = {2026-09-28},
	journal = {Reliability Engineering \& System Safety},
	author = {Costa, Nahuel and Sánchez, Luciano},
	month = jun,
	year = {2022},
	pages = {108353},
}

@article{nicolaou_identifying_2026,
	title = {Identifying {Anomalous} {DESI} {Galaxy} {Spectra} with a {Variational} {Autoencoder}},
	volume = {547},
	issn = {0035-8711},
	doi = {10.1093/mnras/stag010},
	number = {2},
	urldate = {2026-09-28},
	journal = {Monthly Notices of the Royal Astronomical Society},
	author = {Nicolaou, C and Nathan, R P and Lahav, O and Palmese, A and Saintonge, A and Aguilar, J and Ahlen, S and Prieto, C Allende and Bailey, S and BenZvi, S and Bianchi, D and Brodzeller, A and Brooks, D and Claybaugh, T and de la Macorra, A and Della Costa, J and Dey, Arjun and Doel, P and Forero-Romero, J E and Gaztañaga, E and Gontcho A Gontcho, S and Gutierrez, G and Honscheid, K and Howlett, C and Ishak, M and Kehoe, R and Kirkby, D and Kisner, T and Kremin, A and Lambert, A and Landriau, M and Le Guillou, L and Meisner, A and Miquel, R and Moustakas, J and Nadathur, S and Prada, F and Pérez-Ràfols, I and Rossi, G and Sanchez, E and Schubnell, M and Siudek, M and Sprayberry, D and Tarlé, G and Weaver, B A and Zou, H},
	month = apr,
	year = {2026},
	pages = {stag010},
}

@article{khajehnouri_measuring_2019,
	title = {Measuring electrical properties of mortar and concrete samples using the spectral induced polarization method: laboratory set-up},
	volume = {210},
	issn = {0950-0618},
	shorttitle = {Measuring electrical properties of mortar and concrete samples using the spectral induced polarization method},
	doi = {10.1016/j.conbuildmat.2019.03.160},
	urldate = {2019-06-19},
	journal = {Construction and Building Materials},
	author = {Khajehnouri, Yasaman and Chouteau, Michel and Rivard, Patrice and Bérubé, Charles L.},
	month = jun,
	year = {2019},
	pages = {1--12},
}

@article{florsch_inversion_2014,
	title = {Inversion of generalized relaxation time distributions with optimized damping parameter},
	volume = {109},
	issn = {0926-9851},
	doi = {10.1016/j.jappgeo.2014.07.013},
	urldate = {2015-03-25},
	journal = {Journal of Applied Geophysics},
	author = {Florsch, Nicolas and Revil, André and Camerlynck, Christian},
	month = oct,
	year = {2014},
	pages = {119--132},
}

@article{keery_markov-chain_2012,
	title = {Markov-chain {Monte} {Carlo} estimation of distributed {Debye} relaxations in spectral induced polarization},
	volume = {77},
	issn = {0016-8033, 1942-2156},
	doi = {10.1190/geo2011-0244.1},
	language = {en},
	number = {2},
	urldate = {2015-03-26},
	journal = {Geophysics},
	author = {Keery, John and Binley, Andrew and Elshenawy, Ahmed and Clifford, Jeremy},
	month = jan,
	year = {2012},
	pages = {E159--E170},
}

@article{weigand_debye_2016,
	title = {Debye decomposition of time-lapse spectral induced polarisation data},
	volume = {86},
	issn = {0098-3004},
	doi = {10.1016/j.cageo.2015.09.021},
	urldate = {2016-05-06},
	journal = {Computers \& Geosciences},
	author = {Weigand, M. and Kemna, A.},
	month = jan,
	year = {2016},
	pages = {34--45},
}

@article{revil_spectral_2014,
	title = {Spectral induced polarization porosimetry},
	volume = {198},
	issn = {0956-540X},
	doi = {10.1093/gji/ggu180},
	number = {2},
	urldate = {2025-05-20},
	journal = {Geophysical Journal International},
	author = {Revil, A. and Florsch, N. and Camerlynck, C.},
	month = aug,
	year = {2014},
	pages = {1016--1033},
}

@article{abdulsamad_complex_2019,
	title = {Complex {Conductivity} of {Graphitic} {Schists} and {Sandstones}},
	volume = {124},
	copyright = {©2019. American Geophysical Union. All Rights Reserved.},
	issn = {2169-9356},
	doi = {10.1029/2019JB017628},
	language = {en},
	number = {8},
	urldate = {2026-06-15},
	journal = {Journal of Geophysical Research: Solid Earth},
	author = {Abdulsamad, F. and Revil, A. and Ghorbani, A. and Toy, V. and Kirilova, M. and Coperey, A. and Duvillard, P.a. and Ménard, G. and Ravanel, L.},
	year = {2019},
	pages = {8223--8249},
}

@article{macnae_physical_2025,
	title = {A physical interpretation of {Cole}–{Cole} equations and their ambiguous time constants for induced polarization models},
	volume = {243},
	issn = {1365-246X},
	doi = {10.1093/gji/ggaf362},
	number = {2},
	urldate = {2026-06-15},
	journal = {Geophysical Journal International},
	author = {Macnae, James},
	month = nov,
	year = {2025},
	pages = {ggaf362},
}

@inproceedings{virtue_better_2017,
	title = {Better than real: {Complex}-valued neural nets for {MRI} fingerprinting},
	shorttitle = {Better than real},
	doi = {10.1109/ICIP.2017.8297024},
	urldate = {2025-03-21},
	booktitle = {2017 {IEEE} {International} {Conference} on {Image} {Processing} ({ICIP})},
	author = {Virtue, Patrick and Yu, Stella X. and Lustig, Michael},
	month = sep,
	year = {2017},
	pages = {3953--3957},
}

@article{berube_bayesian_2023,
	title = {Bayesian inference of petrophysical properties with generative spectral induced polarization models},
	volume = {88},
	issn = {0016-8033},
	doi = {10.1190/geo2022-0495.1},
	number = {3},
	urldate = {2023-05-05},
	journal = {Geophysics},
	author = {Bérubé, Charles L. and Baron, Frédérique},
	month = may,
	year = {2023},
	pages = {E79--E90},
}

@article{misra_interfacial_2016,
	title = {Interfacial polarization of disseminated conductive minerals in absence of redox-active species — {Part} 1: {Mechanistic} model and validation},
	volume = {81},
	issn = {0016-8033},
	shorttitle = {Interfacial polarization of disseminated conductive minerals in absence of redox-active species — {Part} 1},
	doi = {10.1190/geo2015-0346.1},
	language = {en},
	number = {2},
	urldate = {2021-01-19},
	journal = {Geophysics},
	author = {Misra, S. and Torres-Verdín, C. and Revil, A. and Rasmus, J. and Homan, D.},
	month = mar,
	year = {2016},
	pages = {E139--E157},
}

@article{bucker_electrochemical_2018,
	title = {Electrochemical polarization around metallic particles — {Part} 1: {The} role of diffuse-layer and volume-diffusion relaxation},
	volume = {83},
	issn = {0016-8033},
	shorttitle = {Electrochemical polarization around metallic particles — {Part} 1},
	doi = {10.1190/geo2017-0401.1},
	number = {4},
	urldate = {2021-03-11},
	journal = {Geophysics},
	author = {Bücker, Matthias and Orozco, Adrián Flores and Kemna, Andreas},
	month = mar,
	year = {2018},
	pages = {E203--E217},
}

@article{berube_complex-valued_2025,
	title = {Complex-valued neural networks for spectral induced polarization applications},
	volume = {243},
	issn = {1365-246X},
	doi = {10.1093/gji/ggaf348},
	number = {2},
	urldate = {2025-11-24},
	journal = {Geophysical Journal International},
	author = {Bérubé, Charles L and Gagnon, Sébastien and Kenko, E Rachel and Gagnon, Jean-Luc and Nagasingha, Lahiru M A and Ghanati, Reza and Baron, Frédérique},
	month = nov,
	year = {2025},
	pages = {ggaf348},
}

@inproceedings{rybkin_simple_2021,
	series = {Proceedings of {Machine} {Learning} {Research}},
	title = {Simple and {Effective} {VAE} {Training} with {Calibrated} {Decoders}},
	volume = {139},
	booktitle = {Proceedings of the 38th {International} {Conference} on {Machine} {Learning}},
	publisher = {PMLR},
	author = {Rybkin, Oleh and Daniilidis, Kostas and Levine, Sergey},
	editor = {Meila, Marina and Zhang, Tong},
	month = jul,
	year = {2021},
	pages = {9179--9189},
}

@article{berube_bayesian_2017,
	title = {Bayesian inference of spectral induced polarization parameters for laboratory complex resistivity measurements of rocks and soils},
	volume = {105},
	copyright = {All rights reserved},
	issn = {0098-3004},
	doi = {10.1016/j.cageo.2017.05.001},
	language = {en},
	urldate = {2021-04-21},
	journal = {Computers \& Geosciences},
	author = {Bérubé, Charles L. and Chouteau, Michel and Shamsipour, Pejman and Enkin, Randolph J. and Olivo, Gema R.},
	month = aug,
	year = {2017},
	pages = {51--64},
}

@article{mcinnes_umap:_2018,
	title = {{UMAP}: {Uniform} {Manifold} {Approximation} and {Projection}},
	volume = {3},
	issn = {2475-9066},
	shorttitle = {{UMAP}},
	doi = {10.21105/joss.00861},
	language = {en},
	number = {29},
	urldate = {2019-02-25},
	journal = {Journal of Open Source Software},
	author = {McInnes, Leland and Healy, John and Saul, Nathaniel and Großberger, Lukas},
	month = sep,
	year = {2018},
	pages = {861},
}

@article{nordsiek_new_2008,
	title = {A new approach to fitting induced-polarization spectra},
	volume = {73},
	issn = {0016-8033, 1942-2156},
	doi = {10.1190/1.2987412},
	language = {en},
	number = {6},
	journal = {Geophysics},
	author = {Nordsiek, Sven and Weller, Andreas},
	year = {2008},
	pages = {F235--F245},
}

@incollection{paszke_pytorch_2019,
	title = {{PyTorch}: {An} {Imperative} {Style}, {High}-{Performance} {Deep} {Learning} {Library}},
	booktitle = {Advances in {Neural} {Information} {Processing} {Systems} 32},
	publisher = {Curran Associates, Inc.},
	author = {Paszke, Adam and Gross, Sam and Massa, Francisco and Lerer, Adam and Bradbury, James and Chanan, Gregory and Killeen, Trevor and Lin, Zeming and Gimelshein, Natalia and Antiga, Luca and Desmaison, Alban and Kopf, Andreas and Yang, Edward and DeVito, Zachary and Raison, Martin and Tejani, Alykhan and Chilamkurthy, Sasank and Steiner, Benoit and Fang, Lu and Bai, Junjie and Chintala, Soumith},
	editor = {Wallach, H. and Larochelle, H. and Beygelzimer, A. and Alché-Buc, F. d' and Fox, E. and Garnett, R.},
	year = {2019},
	pages = {8024--8035},
}

@article{weller_relationship_2013,
	title = {On the relationship between induced polarization and surface conductivity: {Implications} for petrophysical interpretation of electrical measurements},
	volume = {78},
	copyright = {© 2013 Society of Exploration Geophysicists},
	issn = {0016-8033, 1942-2156},
	shorttitle = {On the relationship between induced polarization and surface conductivity},
	doi = {10.1190/geo2013-0076.1},
	language = {en},
	number = {5},
	urldate = {2017-11-01},
	journal = {Geophysics},
	author = {Weller, Andreas and Slater, Lee and Nordsiek, Sven},
	month = sep,
	year = {2013},
	pages = {D315--D325},
}

@article{khajehnouri_validation_2020,
	title = {Validation of complex electrical properties of concrete affected by accelerated alkali-silica reaction},
	volume = {113},
	issn = {0958-9465},
	doi = {10.1016/j.cemconcomp.2020.103660},
	language = {en},
	urldate = {2021-04-21},
	journal = {Cement and Concrete Composites},
	author = {Khajehnouri, Yasaman and Rivard, Patrice and Chouteau, Michel and Bérubé, Charles L.},
	month = oct,
	year = {2020},
	pages = {103660},
}

@article{khajehnouri_non-destructive_2020,
	title = {Non-destructive non-invasive assessment of the development of alkali-silica reaction in concrete by spectral induced polarization: {Evaluation} of the complex electrical properties},
	volume = {238},
	copyright = {All rights reserved},
	issn = {0950-0618},
	shorttitle = {Non-destructive non-invasive assessment of the development of alkali-silica reaction in concrete by spectral induced polarization},
	doi = {10.1016/j.conbuildmat.2019.117719},
	language = {en},
	urldate = {2021-04-21},
	journal = {Construction and Building Materials},
	author = {Khajehnouri, Yasaman and Chouteau, Michel and Rivard, Patrice and Bérubé, Charles L.},
	month = mar,
	year = {2020},
	pages = {117719},
}

@mastersthesis{grenon_caracterisation_2018,
	title = {Caractérisation pétrophysique du gisement cuprifère de {Highland} {Valley} ({Colombie}-{Britannique})},
	copyright = {copyright},
	language = {fr},
	urldate = {2025-11-19},
	school = {École Polytechnique de Montréal},
	author = {Grenon, Christophe},
	month = nov,
	year = {2018},
}

@article{berube_mineralogical_2019,
	title = {Mineralogical and textural controls on spectral induced polarization signatures of the {Canadian} {Malartic} gold deposit: {Applications} to mineral exploration},
	volume = {84},
	copyright = {All rights reserved},
	issn = {0016-8033},
	shorttitle = {Mineralogical and textural controls on spectral induced polarization signatures of the {Canadian} {Malartic} gold deposit},
	doi = {10.1190/geo2018-0404.1},
	number = {2},
	urldate = {2021-04-21},
	journal = {Geophysics},
	author = {Bérubé, Charles L. and Olivo, Gema R. and Chouteau, Michel and Perrouty, Stéphane},
	year = {2019},
	pages = {B135--B151},
}

@article{weller_estimating_2010,
	title = {Estimating permeability of sandstone samples by nuclear magnetic resonance and spectral-induced polarization},
	volume = {75},
	issn = {0016-8033, 1942-2156},
	doi = {10.1190/1.3507304},
	language = {en},
	number = {6},
	urldate = {2015-03-17},
	journal = {Geophysics},
	author = {Weller, Andreas and Nordsiek, Sven and Debschütz, Wolfgang},
	month = jan,
	year = {2010},
	pages = {E215--E226},
}

@article{weller_estimation_2010,
	title = {On the estimation of specific surface per unit pore volume from induced polarization: {A} robust empirical relation fits multiple data sets},
	volume = {75},
	issn = {0016-8033},
	doi = {10.1190/1.3471577},
	number = {4},
	urldate = {2025-11-19},
	journal = {Geophysics},
	author = {Weller, Andreas and Slater, Lee and Nordsiek, Sven and Ntarlagiannis, Dimitrios},
	month = jul,
	year = {2010},
	pages = {WA105--WA112},
}

@article{zisser_dependence_2010,
	title = {Dependence of spectral-induced polarization response of sandstone on temperature and its relevance to permeability estimation},
	volume = {115},
	issn = {0148-0227},
	doi = {10.1029/2010JB007526},
	number = {B9},
	urldate = {2025-11-19},
	journal = {Journal of Geophysical Research: Solid Earth},
	author = {Zisser, N. and Kemna, A. and Nover, G.},
	month = sep,
	year = {2010},
}

@article{ustra_spectral_2012,
	title = {Spectral {Induced} {Polarization} ({SIP}) signatures of clayey soils containing toluene},
	volume = {10},
	issn = {1873-0604},
	doi = {https://doi.org/10.3997/1873-0604.2012015},
	number = {6},
	journal = {Near Surface Geophysics},
	author = {Ustra, Andrea and Slater, Lee and Ntarlagiannis, Dimitrios and Elis, Vagner},
	year = {2012},
	pages = {503--515},
}

@article{flores_orozco_delineation_2012,
	title = {Delineation of subsurface hydrocarbon contamination at a former hydrogenation plant using spectral induced polarization imaging},
	volume = {136-137},
	issn = {0169-7722},
	doi = {10.1016/j.jconhyd.2012.06.001},
	journal = {Journal of Contaminant Hydrology},
	author = {Flores Orozco, Adrián and Kemna, Andreas and Oberdörster, Christoph and Zschornack, Ludwig and Leven, Carsten and Dietrich, Peter and Weiss, Holger},
	month = aug,
	year = {2012},
	pages = {131--144},
}

@article{attwa_spectral_2013,
	title = {Spectral induced polarization measurements for predicting the hydraulic conductivity in sandy aquifers},
	volume = {17},
	doi = {10.5194/hess-17-4079-2013},
	number = {10},
	journal = {Hydrology and Earth System Sciences},
	author = {Attwa, M. and Günther, T.},
	year = {2013},
	pages = {4079--4094},
}

@article{weller_prediction_2026,
	title = {On {Prediction} of {Permeability} in {Sedimentary} {Rocks} {Using} {Geoelectrical} {Parameters}},
	volume = {62},
	copyright = {© 2026. The Author(s).},
	issn = {1944-7973},
	doi = {10.1029/2025WR041509},
	language = {en},
	number = {3},
	urldate = {2026-06-12},
	journal = {Water Resources Research},
	author = {Weller, Andreas and Peshtani, Klaudio and Slater, Lee},
	year = {2026},
	pages = {e2025WR041509},
}

@article{peshtani_permeability_2024,
	title = {Permeability and {Induced} {Polarization} of {Mudstones}},
	volume = {60},
	copyright = {© 2024. The Author(s).},
	issn = {1944-7973},
	doi = {10.1029/2024WR037455},
	language = {en},
	number = {8},
	urldate = {2026-06-12},
	journal = {Water Resources Research},
	author = {Peshtani, Klaudio and Weller, Andreas and Slater, Lee},
	year = {2024},
	pages = {e2024WR037455},
}

@article{johansson_spectral_2020,
	title = {Spectral induced polarization of limestones: time domain field data, frequency domain laboratory data and physicochemical rock properties},
	volume = {220},
	issn = {0956-540X},
	shorttitle = {Spectral induced polarization of limestones},
	doi = {10.1093/gji/ggz504},
	number = {2},
	urldate = {2026-06-12},
	journal = {Geophysical Journal International},
	author = {Johansson, Sara and Lindskog, Anders and Fiandaca, Gianluca and Dahlin, Torleif},
	month = feb,
	year = {2020},
	pages = {928--950},
}

@article{revil_comment_2026,
	title = {Comment on ‘{A} physical interpretation of {Cole}–{Cole} equations and their ambiguous time constants for induced polarization models’ by {James} {Macnae}},
	volume = {245},
	issn = {1365-246X},
	doi = {10.1093/gji/ggag003},
	number = {1},
	urldate = {2026-06-12},
	journal = {Geophysical Journal International},
	author = {Revil, A},
	month = apr,
	year = {2026},
	pages = {ggag003},
}

@article{weigand_relationship_2016,
	title = {Relationship between {Cole}–{Cole} model parameters and spectral decomposition parameters derived from {SIP} data},
	volume = {205},
	issn = {0956-540X},
	doi = {10.1093/gji/ggw099},
	number = {3},
	urldate = {2026-06-12},
	journal = {Geophysical Journal International},
	author = {Weigand, M. and Kemna, A.},
	month = jun,
	year = {2016},
	pages = {1414--1419},
}

@article{gurin_time_2013,
	title = {Time domain spectral induced polarization of disseminated electronic conductors: {Laboratory} data analysis through the {Debye} decomposition approach},
	volume = {98},
	issn = {0926-9851},
	doi = {10.1016/j.jappgeo.2013.07.008},
	journal = {Journal of Applied Geophysics},
	author = {Gurin, Grigory and Tarasov, Andrey and Ilyin, Yuri and Titov, Konstantin},
	month = nov,
	year = {2013},
	pages = {44--53},
}

@article{bairlein_influence_2014,
	title = {The influence on sample preparation on spectral induced polarization of unconsolidated sediments},
	volume = {12},
	issn = {1873-0604},
	doi = {https://doi.org/10.3997/1873-0604.2014023},
	number = {5},
	journal = {Near Surface Geophysics},
	author = {Bairlein, K. and Hördt, A. and Nordsiek, S.},
	year = {2014},
	pages = {667--678},
}

@article{zorin_spectral_2015,
	title = {Spectral induced polarization of low and moderately polarizable buried objects},
	volume = {80},
	issn = {0016-8033},
	doi = {10.1190/geo2014-0154.1},
	number = {5},
	urldate = {2025-11-19},
	journal = {Geophysics},
	author = {Zorin, Nikita},
	month = sep,
	year = {2015},
	pages = {E267--E276},
}

@article{picinbono_second-order_1996,
	title = {Second-order complex random vectors and normal distributions},
	volume = {44},
	issn = {1941-0476},
	doi = {10.1109/78.539051},
	number = {10},
	urldate = {2025-12-16},
	journal = {IEEE Transactions on Signal Processing},
	author = {Picinbono, B.},
	month = oct,
	year = {1996},
	pages = {2637--2640},
}

@article{gurin_application_2015,
	title = {Application of the {Debye} decomposition approach to analysis of induced-polarization profiling data ({Julietta} gold-silver deposit, {Magadan} {Region})},
	volume = {56},
	issn = {1068-7971},
	doi = {10.1016/j.rgg.2015.11.008},
	number = {12},
	urldate = {2025-11-20},
	journal = {Russian Geology and Geophysics},
	author = {Gurin, G.V. and Tarasov, A.V. and Il’in, Yu.T. and Titov, K.V.},
	month = dec,
	year = {2015},
	pages = {1757--1771},
}

@article{ustra_relaxation_2016,
	title = {Relaxation time distribution obtained from a {Debye} decomposition of spectral induced polarization data},
	volume = {81},
	issn = {0016-8033},
	doi = {10.1190/geo2015-0095.1},
	number = {2},
	urldate = {2025-11-19},
	journal = {Geophysics},
	author = {Ustra, Andrea and Mendonça, Carlos Alberto and Ntarlagiannis, Dimitrios and Slater, Lee D.},
	month = mar,
	year = {2016},
	pages = {E129--E138},
}

@inproceedings{sohn_learning_2015,
	title = {Learning {Structured} {Output} {Representation} using {Deep} {Conditional} {Generative} {Models}},
	volume = {28},
	urldate = {2021-11-17},
	booktitle = {Advances in {Neural} {Information} {Processing} {Systems}},
	publisher = {Curran Associates, Inc.},
	author = {Sohn, Kihyuk and Lee, Honglak and Yan, Xinchen},
	year = {2015},
    pages = {3483--3491},
}

@article{seigel_mathematical_1959,
	title = {Mathematical formulation and type curves for induced polarization},
	volume = {24},
	issn = {0016-8033, 1942-2156},
	language = {en},
	number = {3},
	urldate = {2014-08-13},
	journal = {Geophysics},
	author = {Seigel, Harold O.},
	month = jan,
	year = {1959},
	pages = {547--565},
}

@article{martin_desaturation_2021,
	title = {Desaturation effects of pyrite–sand mixtures on induced polarization signals},
	volume = {228},
	issn = {0956-540X},
	doi = {10.1093/gji/ggab333},
	number = {1},
	urldate = {2021-11-04},
	journal = {Geophysical Journal International},
	author = {Martin, Tina and Weller, Andreas and Behling, Laura},
	month = jan,
	year = {2021},
	pages = {275--290},
}

@article{zibulski_influence_2023,
	title = {Influence of {Inner} {Surface} {Roughness} on the {Spectral} {Induced} {Polarization} {Response}—{A} {Numerical} {Study}},
	volume = {128},
	copyright = {© 2023. The Authors.},
	issn = {2169-9356},
	doi = {10.1029/2022JB025548},
	language = {en},
	number = {6},
	urldate = {2024-09-04},
	journal = {Journal of Geophysical Research: Solid Earth},
	author = {Zibulski, E. and Klitzsch, N.},
	year = {2023},
	pages = {e2022JB025548},
}

@article{morgan_inversion_1994,
	title = {Inversion for dielectric relaxation spectra},
	volume = {100},
	issn = {0021-9606, 1089-7690},
	doi = {10.1063/1.466932},
	number = {1},
	urldate = {2014-07-25},
	journal = {The Journal of Chemical Physics},
	author = {Morgan, Frank Dale and Lesmes, David P.},
	month = jan,
	year = {1994},
	pages = {671--681},
}

@mastersthesis{benzetta_caracterisation_2024,
	title = {Caractérisation géoélectrique du graphite naturel et application à l'exploration minérale},
	copyright = {copyright},
	language = {fr},
	urldate = {2025-11-28},
	school = {Polytechnique Montréal},
	author = {Benzetta, Rihana},
	month = may,
	year = {2024},
}

@article{hase_conversion_2023,
	title = {Conversion of {Induced} {Polarization} {Data} and {Their} {Uncertainty} from {Time} {Domain} to {Frequency} {Domain} {Using} {Debye} {Decomposition}},
	volume = {13},
	issn = {2075-163X},
	doi = {10.3390/min13070955},
	number = {7},
	journal = {Minerals},
	author = {Hase, Joost and Gurin, Grigory and Titov, Konstantin and Kemna, Andreas},
	year = {2023},
}

@article{wong_electrochemical_1979,
    title = {An electrochemical model of the induced-polarization phenomenon in disseminated sulfide ores},
    volume = {44},
    issn = {0016-8033, 1942-2156},
    doi = {10.1190/1.1441005},
    language = {en},
    number = {7},
    urldate = {2014-07-25},
    journal = {Geophysics},
    author = {Wong, J.},
    month = jul,
    year = {1979},
    pages = {1245--1265},
}
\end{document}